\documentclass[lettersize,journal]{IEEEtran}
\usepackage{array}
\newcommand{\PreserveBackslash}[1]{\let\temp=\\#1\let\\=\temp}
\newcolumntype{C}[1]{>{\PreserveBackslash\centering}p{#1}}
\newcolumntype{R}[1]{>{\PreserveBackslash\raggedleft}p{#1}}
\newcolumntype{L}[1]{>{\PreserveBackslash\raggedright}p{#1}}
\usepackage{etoolbox}
\makeatletter
\patchcmd{\@makecaption}
  {\scshape}
  {}
  {}
  {}
\makeatother
\usepackage{amsmath,amsfonts}
\usepackage{algorithmic}
\usepackage{array}
\usepackage[colorlinks,
            linkcolor=black,       
            anchorcolor=black,  
            citecolor=black,   
            urlcolor = blue,
            ]{hyperref}
\usepackage{textcomp}
\usepackage{bbding} 
\usepackage{stfloats}
\usepackage{url}
\usepackage{verbatim}
\usepackage{graphicx}
\usepackage{subfigure}
\usepackage{color}
\usepackage{textcomp}
\def\BibTeX{{\rm B\kern-.05em{\sc i\kern-.025em b}\kern-.08em
    T\kern-.1667em\lower.7ex\hbox{E}\kern-.125emX}}
\makeatletter
\renewcommand{\@thesubfigure}{\hskip\subfiglabelskip}
\makeatother
\usepackage{balance}
\begin{document}
\title{Unmixing based PAN guided fusion network for hyperspectral imagery}
\author{Shuangliang Li, Yugang Tian*, Hao Xia, Qingwei Liu}


\maketitle

\begin{abstract}
The hyperspectral image (HSI) has been widely used in many applications due to its fruitful spectral information. However, the limitation of imaging sensors has reduced its spatial resolution that causes detail loss. One solution is to fuse the low spatial resolution hyperspectral image (LR-HSI) and the panchromatic image (PAN) with inverse features to get the high-resolution hyperspectral image (HR-HSI). Most of the existing fusion methods just focus on small fusion ratios like 4 or 6, which might be impractical for some large ratios' HSI and PAN image pairs. Moreover, the ill-posedness of restoring detail information in HSI with hundreds of bands from PAN image with only one band has not been solved effectively, especially under large fusion ratios. Therefore, a lightweight unmixing-based pan-guided fusion network (Pgnet) is proposed to mitigate this ill-posedness and improve the fusion performance significantly. Note that the fusion process of the proposed network is under the projected low-dimensional abundance subspace with an extremely large fusion ratio of 16. Furthermore, based on the linear and nonlinear relationships between the PAN intensity and abundance, an interpretable PAN detail inject network (PDIN) is designed to inject the PAN details into the abundance feature efficiently. Comprehensive experiments on simulated and real datasets demonstrate the superiority and generality of our method over several state-of-the-art (SOTA) methods qualitatively and quantitatively (The codes in pytorch and paddle versions
and dataset could be available at \href{https://github.com/rs-lsl/Pgnet}{https://github.com/rs-lsl/Pgnet}). \textcolor{red}{This is a improved version compared with the publication in Tgrs with the modification in the deduction of the PDIN block.} 
\end{abstract}

\begin{IEEEkeywords}
extreme hyperpansharpening; detail injection; hyperspectral unmixing; pan-guided network; multi-scale fusion
\end{IEEEkeywords}

\section{Introduction}
\IEEEPARstart {T}HE hyperspectral image (HSI) provide a rich and distinctive spectral characteristics for different land cover types. Valuable spectral information in HSI has made its applications much widely, such as classification \cite{ref1}-\cite{ref4}, mineral exploitation \cite{ref6}, spectral unmixing \cite{ref7}-\cite{refZheng0}, intelligent agriculture \cite{ref9}, and so on. 

Despite the abundant spectral features it owned from the high spectral resolution, HSI still suffers from the mixed pixel in coarse spatial resolution due to the limitation of imaging sensors, including signal noise ratio (SNR) and satellite revisits period \cite{ref10}. In the aspect of SNR, the narrow wavelength width of HSI bands make the Instantaneous Field of View (IFOV) of the imaging sensor must be designed wider enough to guarantee sufficient energy acception to result in the high SNR. However, the increase of IFOV would enlarge the pixel size and lead to the loss of detail information. 

The economical way to mitigate this limitation is fusing multi-modal images to couple merits of each image and output fused HR-HSI with high spatial and spectral resolution. The common fusion strategies include fusing with multi-spectral (MS) or PAN image. These two types of fusion methods have been explored greatly in recent years especially with the development of deep learning, and they have all achieved excellent fusion performance. Although the MS image could provide spatial and several bands' spectral information into the HSI \cite{refweiwei}, the PAN image with higher spatial resolution than MS image could improve the fused image's spatial detail quality more significantly  \cite{ref11}. Moreover, fusing the LR-HSI and PAN image (termed hyperpansharpening) is more difficult and challenging for the limited single-band's spectral information involved in the latter \cite{refXie}.

Common hyperpansharpening methods could be classified into component substitute (CS)-based, multi-resolution analysis (MRA)-based, variational optimization (VO) and deep learning (DL)-based. CS-based methods generally substitute the simulated intensity band of HSI with the histogram-matched PAN image and then convert them back into original optical image space, mainly including intensity-hue-saturation (IHS) \cite{ref13}, Gram–Schmidt adaptive (GSA) \cite{ref14}\cite{ref15}, principal component analysis (PCA) \cite{ref16}\cite{ref17}. However, these methods usually incur spectral distortion due to the mismatch between the PAN feature and the simulated intensity band. MRA-based methods decompose the PAN image into different spatial scales and inject high-frequency details into the LR-HSI. Widely used methods in this class include smoothing filter-based intensity modulation (SFIM) \cite{ref18}, wavelet transform (Wavelet) \cite{ref19}\cite{ref20}, modulation transfer function with generalized Laplacian pyramid (MTF\_GLP) \cite{ref21}, and MTF-GLP with high-pass modulation (MTF\_GLP\_HPM) \cite{ref22}. However, multi-scale features generated by these methods may lead to spatial distortion. VO-based methods usually formulate the constrained target function with the specific prior and alternatively optimize the variables to obtain the best fusion performance. Among them, coupled nonnegative matrix factorization (CNMF) \cite{ref23} firstly projects the LR-HSI into a subspace, then upsamples the corresponding subspace feature and reconstructs the HR-HSI. In addition, Bayesian sparsity promoted Gaussian prior (Bayesian Sparse) \cite{ref24}, HySure \cite{ref25} and Bayesian naive Gaussian prior (Bayesian Naive) \cite{ref26} are all belong to VO-based methods. Nevertheless, these VO-based methods suffer from the great computation cost \cite{ref10}.

Recently, DL-based fusion methods have demonstrated their superiority and applicability over traditional methods in terms of accuracy and efficiency, especially in image classification and fusion fields \cite{ref2}\cite{ref4}\cite{ref10}. Due to the existence of infinite possible HR-HSI results generated from less informative LR-HSI and PAN images, the hyperpansharpening task is a highly ill-posed problem. Fortunately, DL-based methods are remarkably suitable for this restoration task for their powerful nonlinear learning ability of spectral data distribution and mapping patterns. Especially, the convolutional neural network (CNN) could completely capture the spatial detail distribution property across different regions and has become a vital image process unit. For example, He et al. \cite{ref27}\cite{ref28} proposed the Hyperpnn and HSpeNet structures, which concatenate convoluted LR-HSI and PAN image as input and output fused HR-HSI by cascade convolution layers and reconstruction block with the addition of several short connections. In \cite{ref29}, deep image prior (DIP) is employed to generate upsampled LR-HSI, and the fused results are obtained through cascade spatial and spectral attention blocks. Furthermore, Wele et al. \cite{ref31} Combined the DIP and ‘U-net’ structure to learn detail information in the spatial-expanded feature subspace and achieved promising fusion performance. But learning details in an enlarged feature map is computation costly. 

Actually, most traditional or DL-based hyperpansharpening methods are borrowed by the MS pansharpening field, which may not be appropriate for hyperpansharpening to a certain degree. The number of bands and resolution ratio between LR-HSI and PAN image has increased dramatically in hyperpansharpening task than MS pansharpening. For example, the ZY-1 02D satellite capture the LR-HSI of pixel size 30m with 166 bands and PAN image of spatial resolution 2.5m \cite{ref32}. It means that in the fusion task between these two images, we need to predict 166 bands' spectral value from the PAN image with only one band in each pixel of the fused HR-HSI. Furthermore, the resolution ratio between them is 12 and has never been explored in past hyperpansharpening tasks as far as we know. Most researchers just adopted a ratio of 4 or 6 in simulated experiments \cite{ref27}\cite{ref29}\cite{ref31}, whereas not the entire case in existing remote sensed data and lack of practicability. Consequently, the ill-posedness of the hyperpansharpening task that restoring delicate spectral and spatial information in HR-HSI has become much severe, especially under a large fusion ratio.

To effectively mitigate the above ill-posed problem of restoring detailed information in HR-HSI, we propose a lightweight pan-guided fusion network (Pgnet). Utilizing the relationships between the encoded HSI feature—abundance and PAN intensity, we design an interpretable PAN detail inject network (PDIN) to inject PAN details more effectively. All operations are carried out in the low-dimensional subspace, which could reduce the memory cost greatly. The main parts of Pgnet consist of the abundance encoder, PAN feature guided two-step upsampling and pixel-wise attention blocks. Finally is the HR-HSI reconstruction block. Experiments on different HSI datasets verify the effectiveness and generality of Pgnet in improving the fusion performance quantitatively and qualitatively. The main contributions of this paper could be summarized as follows:

(1) We originally test the hyperpansharpening task in the large fusion ratio of 16 and propose a novel fusion network in low-dimensional abundance subspace to ease this pretty ill-posed image restoration task.

(2) Inspired by the linear and nonlinear relationships between the PAN intensity and abundance feature, an interpretable PAN detail inject network is designed to adjust the second-order statistics of the abundance feature and inject the PAN details. The analysis of the intermediate results verifies the effectiveness of the proposed network.  

(3) Without channel or spatial attention mechanism, a pixel-wise attention block is adopted to reuse the PAN feature and improve the representation accuracy of the abundance feature.

The remaining part of this paper is organized as follows. Section  \uppercase\expandafter{\romannumeral2} introduces the related work with the proposed network and research issues, including auto-encoder, pan detail inject methods and attention mechanism. Section  \uppercase\expandafter{\romannumeral3} describes the methods and detailed network structure. Results of comparative experiments are given in section  \uppercase\expandafter{\romannumeral4}. Discussions about network parts and parameters setting are in section  \uppercase\expandafter{\romannumeral5}. Finally, conclusion are made in section  \uppercase\expandafter{\romannumeral6}.

\section{RELATED WORKS}

\noindent\emph{A.	Auto-encoder} 
\\\indent The original version of auto-encoder aims at designing a combined encoder and decoder network to generate the output image as same as the input image through network optimization \cite{ref39}. Convolution sparse coding (CSC) adopts this structure and designs the super-resolution network \cite{ref40}. The main procedures of CSC include encoding the low-resolution image into a sparse feature map, then upsampling the feature map and decoding it to the super-resolved image. Encoder and decoder in CSC are simulated by convolution blocks with different kernel sizes and channels. Differently, the CSC framework projects the input image into a high-dimensional feature subspace while our designed network is in the low-dimensional subspace with less complexity. This low-dimensional subspace projection is enlightened by the low-rank prior in the hyperspectral image process field \cite{ref42}\cite{refPeng1}.\\

\noindent\emph{B.	Pan Detail Inject Methods}
\\\indent Most of the pan detail inject methods in the hyperpansharpening field are actually derive from the classical MS image pansharpening field, like PAN image decomposition-based \cite{ref18}-\cite{ref22} and guided filter-based \cite{refhe}\cite{refmeng}. For example, Meng et al. \cite{refmeng} proposed an edge-preserving guided filter-based pansharpening method that decomposes the PAN image into three layers include the edge, detail, and low-frequency layer to inject the PAN details. Zhang et al. \cite{refZhang2} proposed a two-stage guided filtering-based MS image pansharpening method to restore the spectral and spatial details in two stages separately. However, these traditional detail injection methods may not be suitable for hyperspectral image with the increment of spectral bands. \\
\indent With the explosive development of deep learning, many CNN-based fusion networks have been designed to inject the pan details into the LR-HSI. For example, He et al. \cite{ref28} generated the PAN details and HSI features through convolution operations and fused these features by cascade CNN blocks to get the HR-HSI. Li et al. \cite{refli} combined the CNN and guided filter to inject the PAN details into the LR-HSI. Dong et al. \cite{refdong} adopted the generative adversarial network to guide the generator learning the distribution of HR-HSI by adversarial training and then inject the PAN details through two representative pansharpening methods with the estimated injection gain. Guan et al. \cite{refguan1} developed a multistage dual-attention guided fusion network to extract the critical information from input images and incorporate these features to get fused HR-HSI.\\
\indent In facing the high memory cost caused by the hundreds of bands in the hyperspectral image, some researchers have tried to utilize the PAN details to enhance the low-dimensional abundance feature generated from the LR-HSI and then multiply by endmembers to get fused HR-HSI \cite{refan}\cite{refwu}. In these works, the enhanced abundance feature is generated by directly adding the extracted high-frequency pan details into the former without logicality. In addition, Lu et al. \cite{ref11} concatenated the intermediately generated MS image and two input images to generate enhanced abundance feature by convolution operations and obtained great fusion performance. Nevertheless, it is hard to pre-define the correlated band groups for different datasets to generate the correct MS image.\\

\noindent\emph{C.	Attention Mechanism} 
\\\indent The typical attention mechanisms in the fusion field include spatial-wise and channel-wise attention. For example, squeeze-and-excitation channel attention \cite{ref43} has been widely used to enhance the informative channels \cite{ref44}. And max-pooling and average-pooling in spatial dimension are employed to adaptively focus on the texture and edge details \cite{ref29}. Based on these attention mechanisms, Zheng et al. \cite{ref29} proposed the dual attention blocks include spectral and spatial attention to improve the quality of the fused HR-HSI. Li et al. \cite{reflijiaojiao} designed a band attention residual block to focus on necessary bands residual information.

In other fields of image enhancement, Hu et al. \cite{ref47} introduced the transformer structure into hyperspectral and multi-spectral fusion tasks which can exploit global relationships of images to improve the fused image quality. Lei et al. \cite{reflei} proposed a nonlocal attention residual network to capture the contextual similarity of all pixels. Dai et al. \cite{refdai} proposed a second-order channel attention model to rescale the channel-wise features by using the second-order statistic feature. Nevertheless, roughly treating all pixels equally across a single band or position in the channel or spatial attention mechanisms is unreasonable, which inevitably incurs the spectral and spatial distortion to results \cite{ref30}. The adopted pixel-wise attention block in this paper could rescale the feature in the entry level to achieve better fusion performance.

\section{METHOD}
This section clarifies the motivation of our network design, including linear and nonlinear relations between PAN intensity and abundance feature. Then detailed network structures including the proposed PAN detail inject network (PDIN) and the entire framework of Pgnet is described. The choice of the loss function is in last.\\

\noindent\emph{A.	Formulation of Fusion Problem} 
\\\indent Ideally, the DL-based hyperpansharpening task in a supervised training manner takes the LR-HSI and PAN images as input and outputs the fused HR-HSI. The loss between the output image and the ground truth would be back propagated to optimize the network parameters iteratively until convergence. However, the real reference HR-HSI is unavailable for the limitation of imaging sensors. And the common method in the real fusion task is to generate LR-HSI and PAN images from the available image pairs by spatial downsampling operations according to Wald protocols \cite{ref51}. Then the original HSI would be regarded as the ground truth. In the simulated experiment, as in Eq.  (\ref{blur_down}) and Eq.  (\ref{srf_down}), the LR-HSI is generated by spatially blurring and downsampling operations from the available HR-HSI, and the PAN image is the linear combination of all HR-HSI bands weighted by the spectral response function (SRF) of the imaging sensor (we denote the images by the capital bold letter and operations by the capital bold italic letter).

\begin{equation}
\label{blur_down}
    \mathbf{Y}=\mathbf{X}BD+\mathbf{N_Y}
\end{equation}
\begin{equation}
\label{srf_down}
    \mathbf{P}=S\mathbf{X}+\mathbf{N_P}
\end{equation}

\noindent where $\textbf{X}\in R^{b\times(mr\times nr)}$ is the HR-HSI, $\textbf{Y}\in R^{b\times(m\times n)}$ means the LR-HSI, $B\in R^{(mr\times nr)\times (mr\times nr)}$ is the blurring kernel matrix, $D\in R^{(mr\times nr)\times (m\times n)}$is the downsampling matrix, $\textbf{P}\in R^{1\times(mr\times nr)}$ represents the PAN image, and $S\in R^{1\times b}$ means the SRF (m and n represent the height and width of $\textbf{Y}$, b is the number of HSI bands, and r is the resolution ratio between $\textbf{Y}$ and $\textbf{X}$). $\mathbf{N_Y}\in R^{b\times(m\times n)}$ and $\mathbf{N_P}\in R^{1\times(mr\times nr)}$ represent the noise terms. Note that we transform these 3D matrices into 2D matrices in band$\times$pixel
manner by integrating the height and width of the image.\\

\noindent\emph{B.	Linear Relation Between Abundance and Pan Intensity}
\indent Endmember and abundance are the critical basis to represent HSI with reduced complexity and retained spectral information \cite{ref8}\cite{ref42}. The endmember is the spectral basis representing the pure pixel, and a pixel's spectra in HSI could be expressed as a linear combination of endmembers weighted by the corresponding abundance fractions as
\begin{equation}
\label{unmixing}
    \mathbf{X}=\mathbf{EA}+\mathbf{N_X}
\end{equation}

\noindent where $\mathbf{X}\in R^{b\times(mr\times nr)}$ indicates the HR-HSI, $\mathbf{E}\in R^{b\times c}$ means endmembers extracted from $\mathbf{X}$ where a column represents an endmember signal, $\mathbf{A}\in R^{c\times(mr\times nr)}$ is the corresponding abundance fractions matrix, and $\mathbf{N_X}\in R^{b\times(mr\times nr)}$ is the noise term. Letter ‘c’ means the endmember number that is much less than the number of HSI bands.  


In fact, the intensity of PAN image is still a linear combination of the abundance feature. In detail, we substitute $\mathbf{X}$ in Eq. (\ref{srf_down}) with Eq. (\ref{unmixing}) and get
\begin{equation}
\label{temp0}
    \mathbf{P}=S\mathbf{EA}+S\mathbf{N_X}+\mathbf{N_P}
\end{equation}

\noindent Then let $S\mathbf{E}$ as $S’$, $S\mathbf{N_X}+\mathbf{N_P}$ as $\mathbf{N’}$ and get
\begin{equation}
\label{linear}
    \mathbf{P}=S’\mathbf{A}+\mathbf{N’}
\end{equation}
where $S’\in R^{1\times c}$ represents the SRF of  PAN image from the abundance feature. Eq. (\ref{linear}) means that the PAN image is a linear combination of the low-dimensional abundance map with the addition of the noise term. Therefore, the injection of PAN details into the abundance feature is comparable to the original HSI in Eq. (\ref{srf_down}). In addition, taking the fusion task in the low-dimensional abundance subspace could release this ill-posed fusion problem greatly and maintain the high correlations among spectral bands \cite{ref53}.\\

\begin{figure}[t]
\centering
    \includegraphics[width=0.5\textwidth,trim=80 20 260 300,clip]{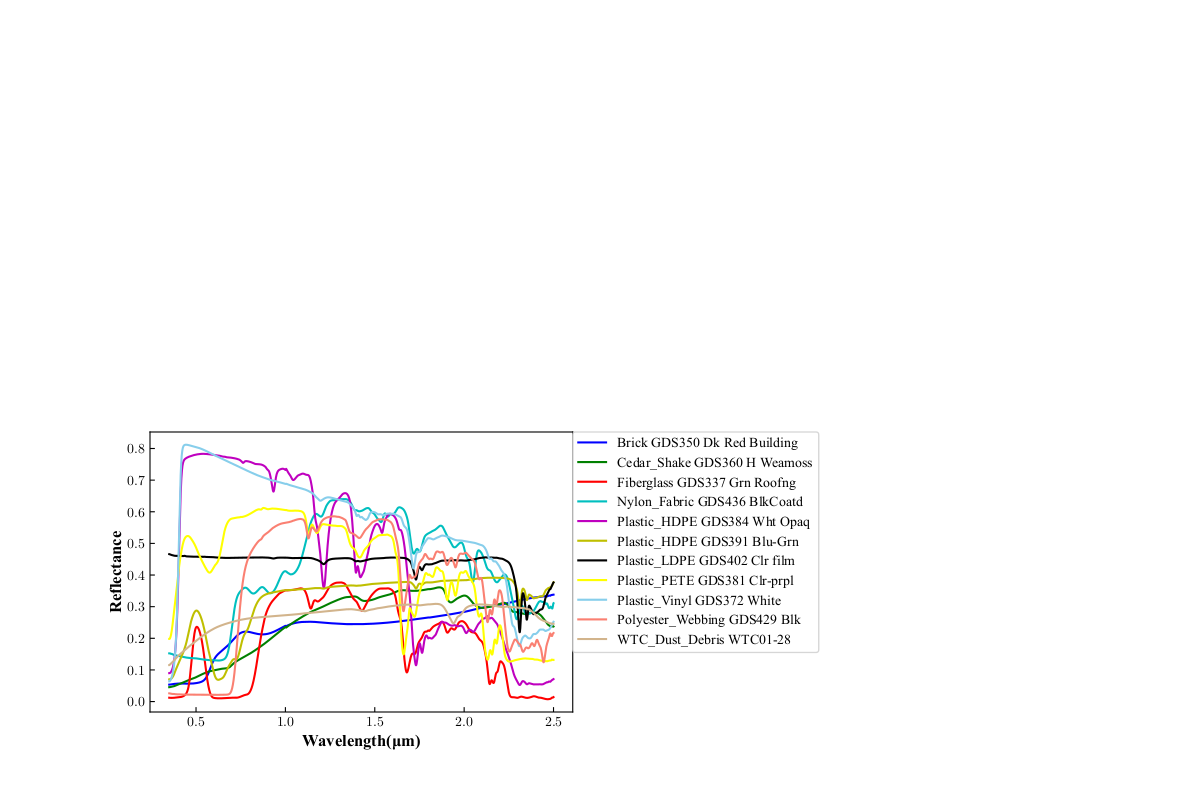}
    \caption{Spectral curve of pure endmember that selected from the USGS spectral library. X-axis means the spectral wavelength from 0 $\mu$m to 2.5 $\mu$m, and y-axis indicates the reflectance.}
    \label{fig:spectral curve}
\end{figure}
\begin{figure}[t]
\centering
    \includegraphics[width=0.45\textwidth,trim=0 0 0 40,clip]{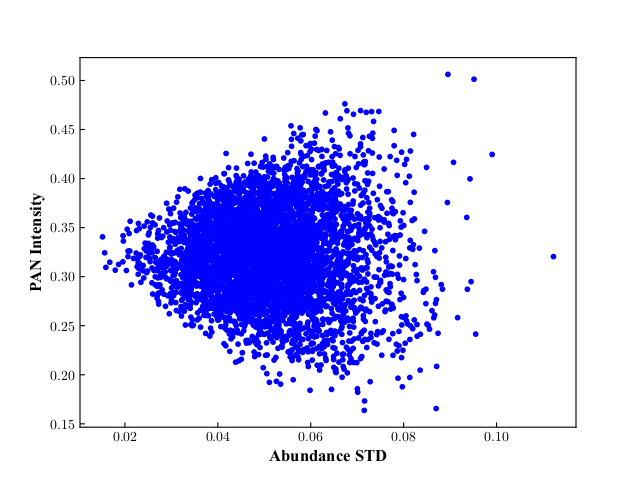}
    \caption{Simulated scatter points between PAN intensity and abundance STD with the ‘fish’ distribution. The x-axis denotes the abundance STD of the simulated abundance pixels along the channel dimension, while the y-axis represents the corresponding pixel's PAN intensity.}
    \label{fig:scatter points}
\end{figure}

\noindent\emph{C.	Nonlinear Relations Between Abundance Feature and Pan Intensity}\\
\indent In addition to the linear relation, we experimentally find the general nonlinear correlation between the PAN intensity and the second-order statistic feature—standard deviation (STD) of the abundance feature along the channel dimension. In this experiment, we randomly choose the endmember spectra from the United States Geological Survey (USGS) spectrum library, consisting of 11 pure materials plotted in Fig. \ref{fig:spectral curve}. The spectral values of these materials fluctuate greatly and could represent the diversity of ground objects well. At the same time, we initialize the abundance map randomly with 5000 pixels under the uniform distribution between 0$-$1 and normalize abundance in the sum-to-one manner for each pixel. Then abundance features are multiplied by selected endmembers to get mixed HSI pixels. Lastly, we generate the PAN image by multiplying the SRF of worldview2 with the simulated hyperspectral image and summing the results channel-wise. 

\begin{figure}[t]
\centering
    \includegraphics[width=0.5\textwidth,trim=310 230 150 270,clip]{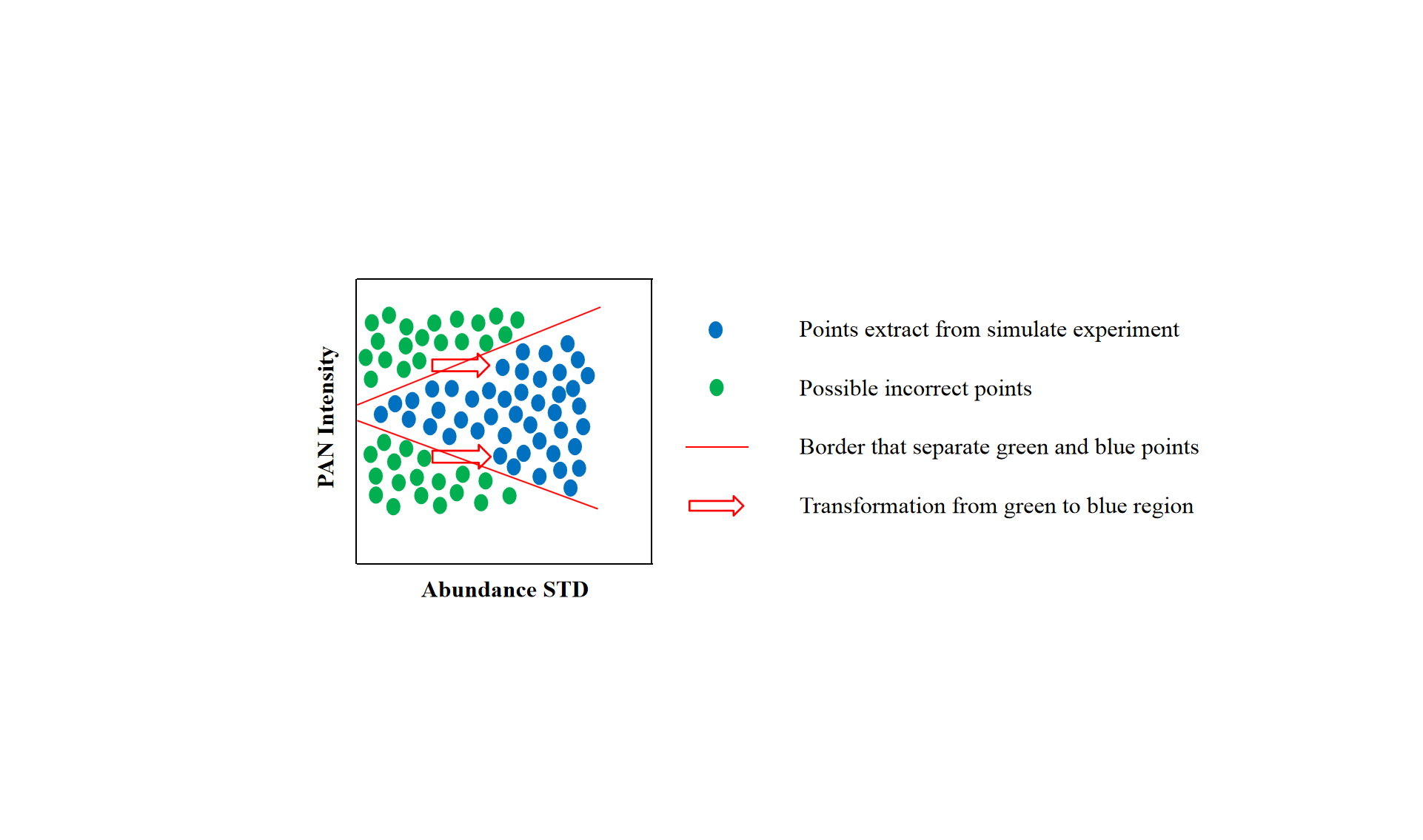}
    \caption{Abstract graph from the Fig. \ref{fig:scatter points}. Blue points represent the points that satisfy the ‘fish’ distribution, while green points are the possible incorrect points that need to transform.}
    \label{fig:abstract fig}
\end{figure}
\begin{figure}[t]
\centering
    \includegraphics[width=0.5\textwidth,trim=145 50 200 10,clip]{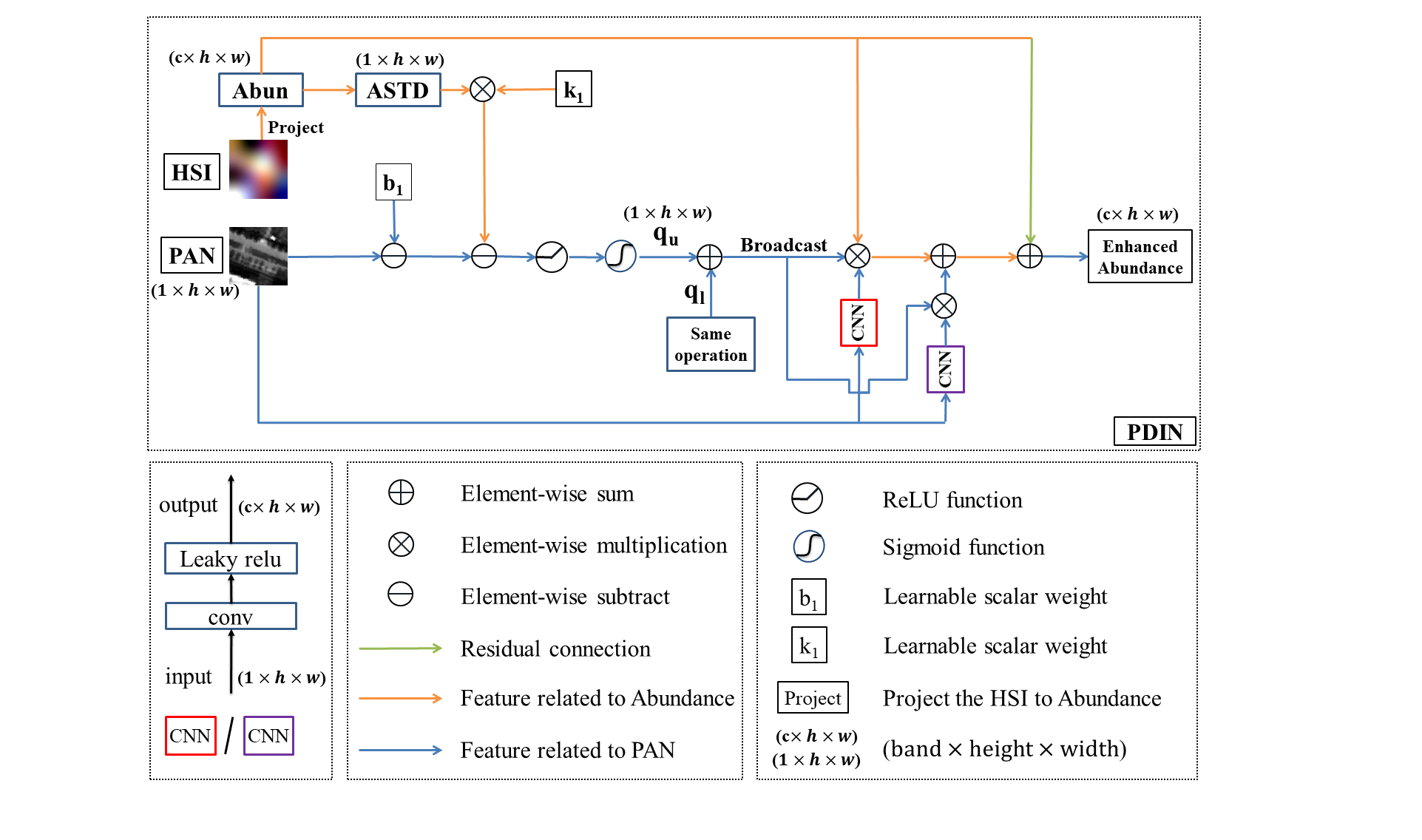}
    \caption{Our proposed pan detail inject network (PDIN). ‘HSI’, ‘PAN’, ‘Abun’ and ‘ASTD’ notations in square blocks mean the hyperspectral image, original or convoluted PAN feature, Abundance feature and Abundance STD. The ‘same operation’ in the block means the ‘$q_l$’ is generated by the same operation as ‘$q_u$’.}
    \label{fig:three inject}
\end{figure}

\indent This nonlinear correlation is illustrated by the distribution of simulated pixels in Fig. \ref{fig:scatter points}, and we name it as ‘fish’ distribution with a sharp head and disperse tail. On the one hand, it shows that the lower STD of abundance, which means the pixel is mixed greatly by many kinds of materials, then the more concentrated of the PAN intensity, approximately 0.33 in Fig. \ref{fig:scatter points}. On the other hand, with the increase of the abundance STD that represents the pureness of a pixel is enlarging, the distribution of the PAN intensity tends to be more dispersed. This nonlinear relationship between PAN intensity and abundance STD and the linear relationship in Eq. (\ref{linear}) have inspired the design of a detail inject network under abundance feature subspace.\\

\noindent\emph{D.	Pan Detail Inject Network}

In this section, based on the linear and nonlinear relationships between PAN intensity and abundance STD (ASTD) described in the section \uppercase\expandafter{\romannumeral3}-B and \uppercase\expandafter{\romannumeral3}-C, we design a novel PAN detail inject network (PDIN) as shown in Fig. \ref{fig:three inject}.

Firstly, we extract the central part of Fig. \ref{fig:scatter points} and redraw it in Fig. \ref{fig:abstract fig}. Blue points represent the samples that obey the ‘fish’ distribution. Green points mean the possible incorrect abundance pixels generated by the specific modules in the network (i.e., upsample module). In order to generate the correct abundance feature, it is required to transform the green points to the correct subspace of ‘fish’ distribution (blue region). And the verification of transform patterns would be detailed in section \uppercase\expandafter{\romannumeral5}-D.   

\indent \textcolor{red}{In fact, the pan intensity in reference pan image play the role of supervision, which means the corresponding abundance pixel's pan intensity would not been changed during transformation process. As shown in Fig. \ref{fig:abstract fig}, the red arrow means the transformation of abundance pixels' STD which is in horizontal direction without change of pan intensity.} The adjustments of abundance STD include two steps. The first is to identify the green points and compute the STD distance they need to move. The second step is to move the green points into the blue region. Detailed derivations are as follows.

(1). Firstly, we need to separate the green points from the blue points using red border lines as shown in Fig. \ref{fig:abstract fig}. We can denote these two red lines with the equations as
\begin{equation}
\label{redline1}
    y={k_1}{x}+b_1
\end{equation}
\begin{equation}
\label{redline2}
    y={k_2}{x}+b_2
\end{equation}
where Eq. \ref{redline1} represents the upper red line, and Eq. \ref{redline2} means the lower red line in Fig. \ref{fig:abstract fig}. $y$ represents the y-axis—PAN intensity and $x$ represents the x-axis—abundance STD. $k_1$, $k_2$ are slopes, and $b_1$, $b_2$ are intercepts of these two red lines, respectively. Note that we set slopes and intercepts the trainable parameters for they are correlated with the data distribution. These parameters are updated by back propagation and the initialization method would be determined in section \uppercase\expandafter{\romannumeral5}-E.

Then according to the linear programming method that the green points meet the following two conditions 
\begin{equation}
\label{aboveredline1}
    {y}-({k_1}{x}+b_1)\textgreater 0
\end{equation}
\begin{equation}
\label{belowredline2}
    {y}-({k_2}{x}+b_2)\textless 0
\end{equation}
It means that the green points satisfy the Eq. \ref{aboveredline1} are of which above the upper red line, and Eq. \ref{belowredline2} holds the opposite meaning. 

In the next step, we compute the weight of distance that green points need to move. It could be inferred from Fig. \ref{fig:abstract fig} that the farther from the red line of green points, the more distance they need to move. So for the green points above the upper red line, we subtract the $y$ by $({k_1}{x}+b_1)$ to measure the moving weight of upper green points and the use of sigmoid function could rescale their weights into $[0, 1]$. This is the same meaning as the green points under the lower red line. Note that we embed the Rectified Linear Unit (ReLU) function to inactivate the moving weights of blue points in this step. 
\begin{equation}
\label{weight1}
    q_u=F_s(F_{R}({y}-({k_1}{x}+b_1)))
\end{equation}
\begin{equation}
\label{weight2}
    q_l=F_s(F_{R}(({k_2}{x}+b_2)-{y}))
\end{equation}
where $q_u$ and $q_l$ are weight maps which measure the distance that the top and bottom green points need to move. $y$ represents all PAN pixels in a patch while $x$ is the corresponding abundance STD pixels. $F_{R}$ represents the ReLU function and $F_s$ means the sigmoid function.

At last, we get the moving weight $q$ of all pixels in a patch by summing the $q_u$ and $q_l$ as follows. Note that this operation could make the weights of blue points equal to 1 and the weights of green points great than 1.
\begin{equation}
\label{weight_all}
    q=q_u+q_l
\end{equation}

(2). After getting the moving weight, we need to transform the green points to the blue region through the change of abundance STD. And this transformation could be completed by multiplying the pixel with the moving weight $q$ as
\begin{equation}
\label{multi_std}
    f_{STD}(q\mathbf{X})=qf_{STD}(\mathbf{X})
\end{equation}
where $f_{STD}$ represents the computation of abundance STD to get ASTD, as shown in Fig. \ref{fig:three inject}. This equation means that multiplying the pixel $\mathbf{X}$ by weight $q$ could change the STD of pixel $\mathbf{X}$ from $f_{STD}(\mathbf{X})$ into $qf_{STD}(\mathbf{X})$.

Additionally, the moving weight in Eq. (\ref{weight_all}) may not significantly improve the detail injection effect for it only considers the distance along the y-axis as in Eq. (\ref{weight1}) and (\ref{weight2}). And the blue points that fit the ‘fish’ distribution may also need a little movement to improve the representation accuracy of the abundance feature further. Therefore, we additionally learn the moving weight from PAN intensity to adjust the moving distance of all pixels, which is inspired by the linear relationships between PAN intensity and abundance in Eq. (\ref{linear}). Due to the fruitful and complex spatial details in the PAN image, the CNN is selected to learn this moving weight, and its network structure is composed of a single convolution layer and Leaky ReLU function, as in the bottom-left part of Fig. \ref{fig:three inject}. This adjustment could be formulated as
\begin{equation}
\label{std_change}
    \mathbf{X_e}=q{P_1}\mathbf{X}
\end{equation}
where $\mathbf{X_e}$ means the enhanced abundance feature, $q$ is from Eq. (\ref{weight_all}) and $P_1$ is the weight learned from the PAN feature. And we multiply the $P_1$ by $q$ to focus on the green points.

\textcolor{red}{However, this multiplication operation would change the PAN intensity of the corresponding pixel which may inconsistent with the reference pan intensity. Therefore, a bias term is added to adjust the PAN intensity of the corresponding abundance pixel further as (the bias term wouldn't change the STD of the abundance pixel)}
\begin{equation}
\label{std_change_last}
    \mathbf{X_e}=q{P_1}\mathbf{X}+qP_2
\end{equation}
where $P_1$ is the multiplicative weight and $P_2$ is the additive weight. They are all learned from the PAN feature by CNN structure. We also multiply the $P_2$ by $q$ to focus on the green points. $\mathbf{X_e}$ represents the completely enhanced abundance feature. Furthermore, we add the residual connection to improve the learning speed and stability, as indicated by the green arrow in Fig. \ref{fig:three inject}. The effectiveness of these terms would be tested in section \uppercase\expandafter{\romannumeral5}-C.\\

\begin{figure*}[t]
\centering
    \includegraphics[width=1\textwidth,trim=70 200 130 200,clip]{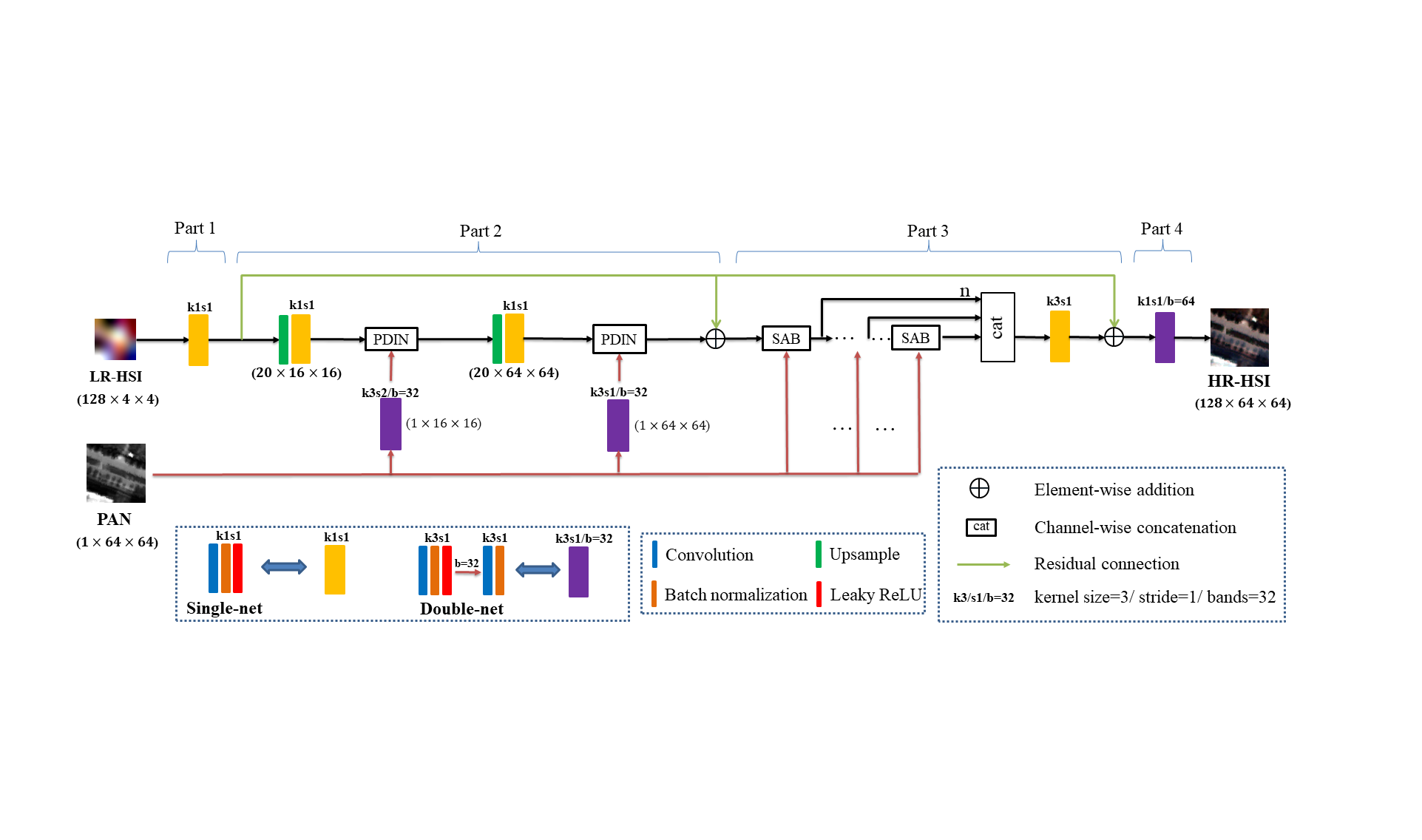}
    \caption{The overall network structure of the proposed Pgnet. ‘LR-HSI’ and ‘PAN’ image are the input of the network, and ‘HR-HSI’ is the fused result. Their sizes are 128×4×4, 1×64×64 and 128×64×64 in the Chikusei dataset, which represent the band number, height and width respectively. Letter ‘b’ above arrows represents band numbers after the last operation. Number ‘n’ before ‘cat’ notation means that there are n features would be concatenated. We denoted the ‘single-net’ by yellow block and ‘double-net’ by purple block. The black arrow indicates the abundance-related feature transmission. The red arrow means the transmission of the PAN feature. The green line means the residual connection of the abundance map upsampled by the ‘bicubic’ interpolation method. We would detail the SAB block in section \uppercase\expandafter{\romannumeral3}-E(3).}
    \label{fig:network}
\end{figure*}

\noindent\emph{E.	 Network Structure }

As depicted in Fig. \ref{fig:network}, the main structures of Pgnet consist of four parts: (1) abundance feature encodes, (2) detail injection-based upsample, (3) pixel-wise attention mechanism, and (4) HR-HSI reconstruction. Two short connections from the end of part 1 are added to the end of part 2 and part 3 to force the network learning high-frequency details, as indicated by green arrows in Fig. \ref{fig:network}. Detailed introductions of the network are in the following.\\

\noindent\emph{1)	Abundance feature encode network}

The first part is to encode the input LR-HSI into the abundance feature. We simulate the abundance feature extractor by a convolution layer followed by the batch normalization and Leaky ReLU activation function. We group these three operations as single-net denoted by yellow block in Fig. \ref{fig:network}, and this encoding process could be expressed as
\begin{equation}
\label{encoding}
    {\mathbf{Y_{abun}}}={f_{sig}}(\mathbf{Y})
\end{equation}

\begin{figure}[t]
\centering
    \includegraphics[width=0.5\textwidth,trim=180 330 700 180,clip]{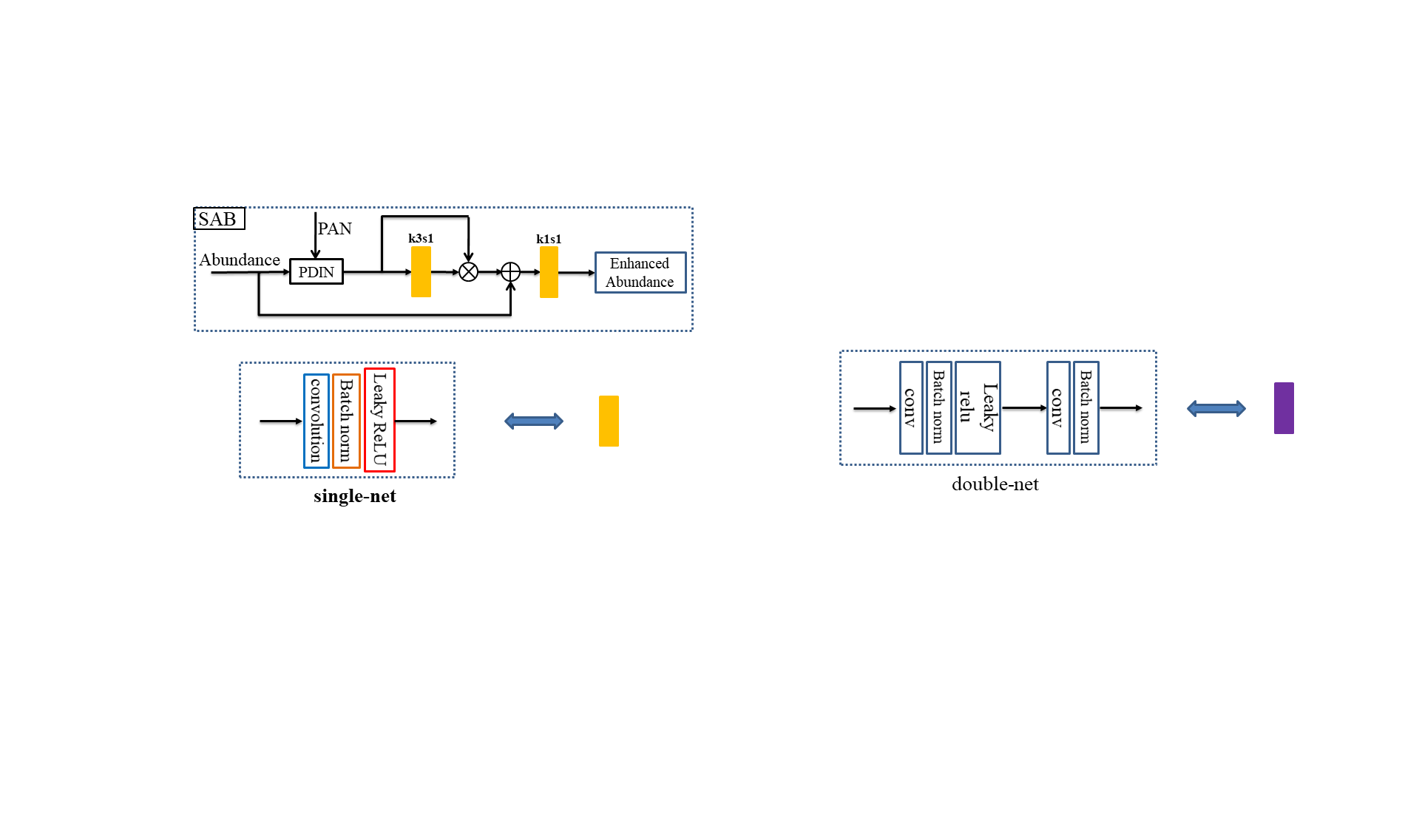}
    \caption{Network structure of SAB in the part 3 of Pgnet. Notations in this figure are the same as Fig. \ref{fig:network}.}
    \label{fig:subnetwork}
\end{figure}

\noindent where $\mathbf{Y_{abun}}$ is the extracted abundance feature map whose channels are set to 20 according to the experiments in section \uppercase\expandafter{\romannumeral5}-A. Kernel size and stride are all set to 1 in this part.

Some works have been devoted to unmixing the HSI to get endmember signal and abundance feature simultaneously by delicately designed VO-based methods \cite{ref7}\cite{refPeng} or DL-based methods \cite{refHong0}\cite{refGao0}. Differently, since we only need to extract the abundance feature from the LR-HSI, we just adopt the single convolution block in this part. Convolution operation in this part is the multiplication between the convolution kernel parameters (represent possible ‘endmember signal’) and hyperspectral image pixel. In other words, a resulted abundance element is the inner product of two vectors within the former two matrices. And the inner product could measure the similarity between two vectors. So this single convolution operation could result in the approximate abundance feature (similar to the Fourier transform to get the Fourier coefficient using the inner product or weighted integral method). Meanwhile, kernel size of 1 could decrease network parameters to relieve computation cost.\\

\noindent\emph{2)	PAN image detail injection-based upsample network }

We decompose 16 times' upsampling of the abundance feature map into two successive 4 ratios' upsample steps indicated by two green blocks in Fig. \ref{fig:network}. As for the upsampling method, we choose the bicubic interpolation method. Then followed by a single-net of kernel size 1 to optimize the feature representation. This combination of operations for upsampling and its effect have been detailed in \cite{ref54}.

Then the downsampled PAN feature with one channel and the upsampled abundance features are fed into the PDIN block. Noticeably, the PAN feature is generated by double cascade single–net (double-net) denoted by $f_{dounet32}$, as shown in the purple block of Fig. \ref{fig:network}, where 32 means the intermediate feature has 32 channels. This part is formulated as 
\begin{equation}
\label{upsample4}
    {\mathbf{Y_{abun4}}}={f_{up}}(\mathbf{Y_{abun}})
\end{equation}
\begin{equation}
\label{down4}
    {\mathbf{P_{fea4}}}={f_{dounet32}}(\mathbf{PAN})
\end{equation}
\begin{equation}
\label{PDIN4}
    {\mathbf{Y_{abun4}}}={f_{PDIN}}(\mathbf{Y_{abun4}}, \mathbf{P_{fea4}})
\end{equation}
where $\mathbf{Y_{abun4}}$ in Eq. (\ref{upsample4}) means four times upsampled abundance through the upsampling module$-f_{up}$, and $\mathbf{P_{fea4}}$ is the downsampled PAN feature with the same resolution as $\mathbf{Y_{abun4}}$. Note that we select the stride convolution to dowmsample the PAN image which could capture the various texture and edge patterns in the PAN image. At last, $f_{PDIN}$ which means the proposed PDIN in section \uppercase\expandafter{\romannumeral3}-D injects the high-frequency information in the convoluted PAN feature into the abundance to get enhanced abundance feature$-\mathbf{Y_{abun4}}$, as in Eq. (\ref{PDIN4}).  

We obtain the $\mathbf{Y_{abun16}}$ which means the 16 times upsampled abundance map as formulated in Eq. (\ref{eq20})-(\ref{eq22}). It is of the same generating operations as the $\mathbf{Y_{abun4}}$ and would enter in part 3.
\begin{equation}
\label{eq20}
    {\mathbf{Y_{abun16}}}=f_{up}(\mathbf{Y_{abun4}})
\end{equation}
\begin{equation}
\label{eq21}
    {\mathbf{P_{fea}}}={f_{dounet32}}(\mathbf{PAN})
\end{equation}
\begin{equation}
\label{eq22}
    {\mathbf{Y_{abun16}}}=f_{PDIN}(\mathbf{Y_{abun16}, P_{fea}})+f_{up16}(\mathbf{Y_{abun}})
\end{equation}
where $f_{up16}$ means the 16 times' umsample operation by bicubic method.\\
\\
\noindent\emph{3)	 Pan-guided pixel-wise attention network}

The fined spectral and spatial details in HR-HSI make the adopted upsample operations difficult to achieve the excellent fusion performance. So we introduce the pixel-wise attention mechanism and utilize the PAN feature again to enhance the representation accuracy of the abundance feature.

Firstly, as shown in Fig. \ref{fig:subnetwork}, in each self-attention block (SAB), the PAN feature and abundance feature are fed into the PDIN block to get partly enhanced abundance feature denoted by $\mathbf{Y_{fea}}$. Then pixel-wise multiplicative weight are generated to get the entirely enhanced abundance feature $\mathbf{Y_{abun16}}$ with the residual connection as
\begin{equation}
\label{eq23}
    {\mathbf{Y_{fea}}}={f_{PDIN}}(\mathbf{Y_{abun16}, PAN})
\end{equation}
\begin{equation}
\label{eq24}
    {\mathbf{Y_{mid}}}=\mathbf{Y_{fea}}*{f_{sig\_m}}(\mathbf{Y_{fea}})
\end{equation}
\begin{equation}
\label{eq25}
    {\mathbf{Y_{abun16}}}={f_{sig}}(\mathbf{Y_{abun16}+Y_{mid}})
\end{equation}
where $*$ means the element-wise multiplication and $f_{sig\_m}$ representing the single-net is used to generate multiplicative weight. Another single-net—$f_{sig}$ is then used to improve the detail restoration further.

Moreover, the enhanced abundance features in different layers denoted by $\mathbf{Y_{abun16\_1}}$ to $\mathbf{Y_{abun16\_4}}$ would be concatenated to reuse the multi-depth information with the single-net, as shown in the part 3 of Fig. \ref{fig:network}. Note that the number of layers (SAB blocks number) is determined experimentally in section \uppercase\expandafter{\romannumeral5}-A and number 4 gets the compromise effect between the accuracy and complexity. Short connection is also added to accelerate convergence as 
\begin{equation}
\label{eq26}
    {\mathbf{Y_{abun\_all}}}={f_{concat}}(\mathbf{Y_{abun16\_1}, ..., Y_{abun16\_4}})
\end{equation}
\begin{equation}
\label{eq27}
    {\mathbf{Y_{abun\_last}}}={f_{sig}}(\mathbf{Y_{abun\_all}})+f_{up16}(\mathbf{Y_{abun}})
\end{equation}
where $f_{concat}$ means the concatenate function and $\mathbf{Y_{abun\_all}}$ represents the concatenated feature. $\mathbf{Y_{abun\_last}}$ is the totally enhanced abundance feature and would enter in the next part.\\

\noindent\emph{4)	HR-HSI reconstruction network }

The double-net is adopted to restore the HR-HSI, as shown in Fig. \ref{fig:network}. These filter kernels and activation function are simulations of the endmembers. Convolution operation in this part is equal to multiplying the abundance feature by endmember signals and obtaining the fused HR-HSI. Also, the nonlinear characteristic of the activation function-Leaky ReLU could improve the reconstruction performance considering the nonlinear assumption in the spectral mixing model \cite{ref8}. 
\begin{equation}
\label{eq28}
    {\mathbf{Y_{HR-HSI}}}={f_{dounet64}}(\mathbf{Y_{abun\_last}})
\end{equation}
where $f_{dounet64}$ means the ‘dounet’ with 64 intermediate channels and $\mathbf{Y_{HR-HSI}}$ is the reconstructed HR-HSI.\\

\noindent\emph{F.	Loss Function}

We choose the MSE loss \cite{ref27}\cite{ref28} to optimize the parameters of the proposed Pgnet for its fast convergence rate as 

\begin{equation}
\label{eq29}
    MSE(\hat{\mathrm{\mathbf{X}}},\ \mathbf{X})
    =
    \frac{1}{N_{m}}\sum_{j}\Vert\hat{\mathrm{x}}_{j}-\mathrm{x}_{j}\Vert_{2}^{2}  
\end{equation}

\noindent where $\hat{\mathrm{\mathbf{X}}}$ means the fused image and $\mathbf{X}$ represents the reference image. $\hat{\mathrm{x}}_{j}$ and ${\mathrm{x}}_{j}$ are the $i^{th}$ pixels from these two images respectively, and $N_m$ is the number of image pixels. In addition, to avoid the blurring effect induced by this loss function, we add the SAM loss to improve the spectral fidelity of the fused HR-HSI as

\begin{equation} 
\label{eq30}
{SAM} (\hat{\mathrm{\mathbf{X}}},\ \mathbf{X})=\frac{1}{N_{m}}\sum_{j}\mathrm{arccos} \frac{\hat{\mathrm{x}}_{j}^{\top}\mathrm{x}_{j}}{\Vert\hat{\mathrm{x}}_{j}\Vert_{2}\Vert \mathrm{x}_{j}\Vert_{2}}
\end{equation}
where $\hat{\mathrm{x}}_{j}^{\top}\mathrm{x}_{j}$ represents the inner product of these two vectors, and the other notations in this equation are the same as Eq. (\ref{eq29}). Then we get the combined loss function to train the network as follows, where $\alpha$ is the weight parameter of SAM loss and it would be determined in section \uppercase\expandafter{\romannumeral5}-E for optimal fusion performance. 
\begin{equation}
\label{eq31}
    Loss(\hat{\mathrm{\mathbf{X}}},\ \mathbf{X})=MSE(\hat{\mathrm{\mathbf{X}}},\ \mathbf{X})+\alpha SAM(\hat{\mathrm{\mathbf{X}}},\ \mathbf{X})
\end{equation}

\section{EXPERIMENTAL RESULTS}
This section shows the comparative experimental results in the simulated and real datasets. At first, We describe the characteristics of these datasets and training details, including hyperparameters setting, compared methods, and fusion performance evaluation indices. Then we present the fused results on four datasets qualitatively and quantitatively. Finally, different scales' fusion tests on the Chikusei dataset are done.\\

\noindent\emph{A. Experimental Datasets }\\
\indent 1). JiaXing dataset: This dataset was acquired by an unmanned aerial vehicle in the developed city JiaXing, China. Dominant land types in this scene include buildings, roads, a few trees and lakes. Its spatial resolution is 1.2m, and spectral wavelength ranges from 393 to 983nm. The size of the entire HR-HSI scene is 126×1072×12944, where 126 is the band number, and the other two are height and width. Following the Wald protocol \cite{ref51}, we generate the simulated LR-HSI by Gauss blurring and downsampling. The kernel size of the gauss filter kernel is equal to 16, and its standard deviation (STD) is set to 0.8493, which are the same as other datasets. The PAN image is generated from the linear combination of HR-HSI weighted by SRF of the worldview 2 satellite sensor. So the sizes of LR-HSI and PAN images are 126×67×809 and 1×1072×12944, respectively.\\
\indent 2). Chikusei dataset \cite{ref55}: The Chikusei scene was captured by the Headwall Hyperspectral-VNIR-C imaging sensor over the agricultural and urban area in Chikusei, Japan. The original Chikusei image consists of 128 spectral bands spanning from 363 to 1018 nm. The spatial size is 2517×2335 with a resolution of 2.5m. By the same operation as the JiaXing dataset, the sizes of generated LR-HSI and PAN images are 128×157×145 and 1×2512×2320, respectively.\\
\indent 3). XiongAn dataset \cite{ref56}: The XiongAn hyperspectral image was captured on Matiwan Village in XiongAn New Area, China. Different kinds of crops and grasses cover this Village mostly. This dataset includes 256 bands with a high spatial resolution of 0.5m. Its spectral wavelength ranges from 391 to 1002nm with a wavelength width of 4.5nm. The image height and width are 1580 and 3750. With the same operation above, LR-HSI and PAN images are of sizes 256×98×234 and 1×1568×3744, respectively.
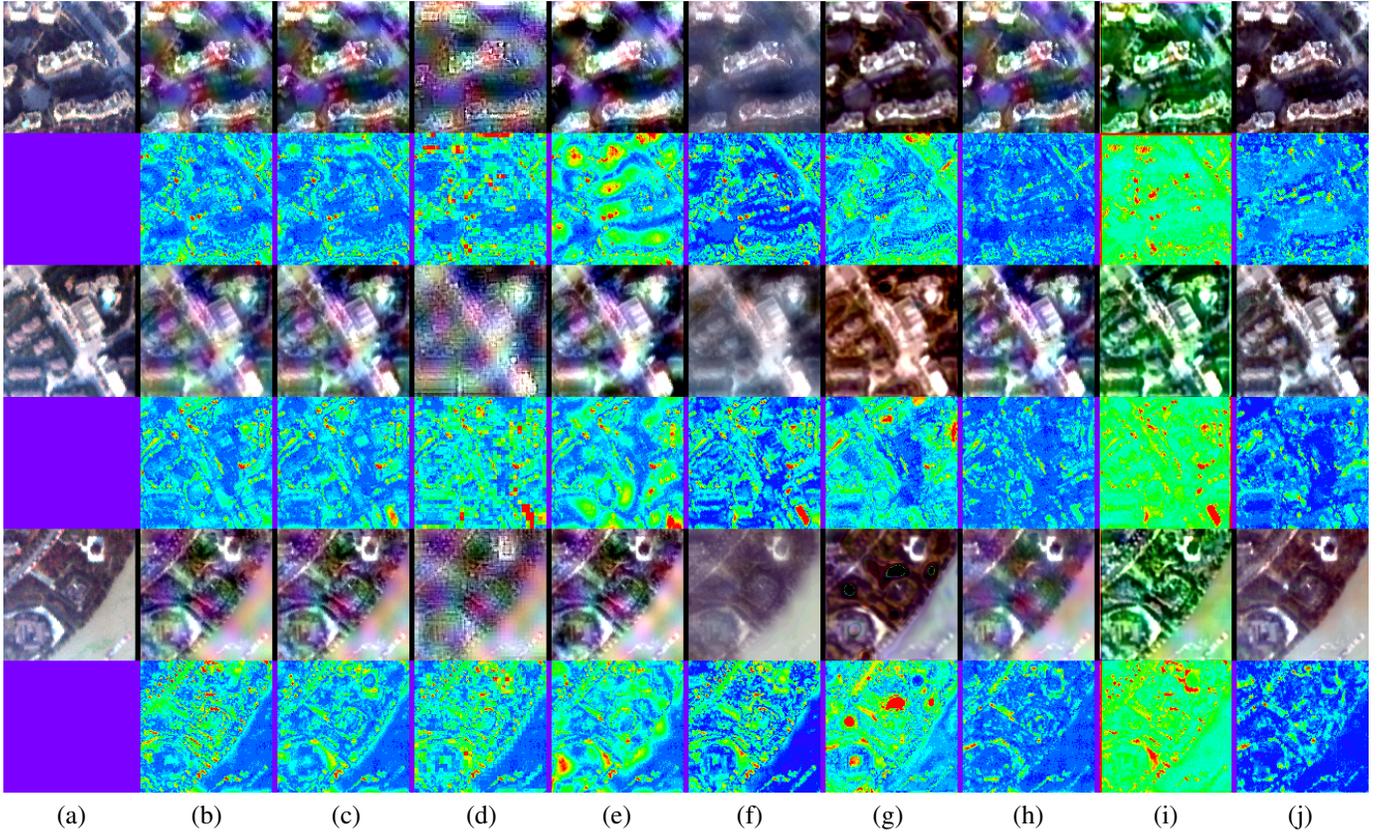
\begin{figure*}[t]
    \centering
    \includegraphics[width=1\textwidth, trim=0 0 0 0, clip]{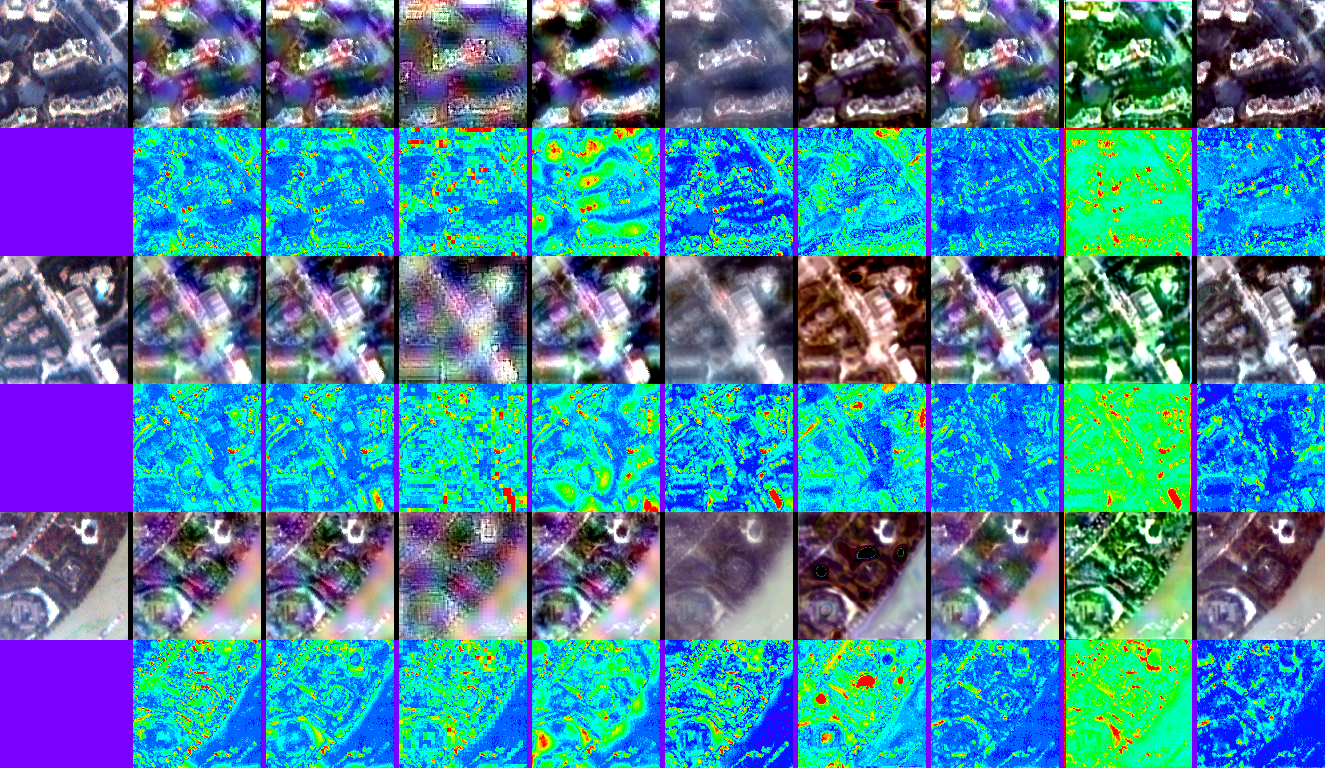}
    
    \begin{picture}(0,0)
    \put(-238,0){(a) \qquad\quad\; (b)  \qquad\quad\;\, (c) \qquad\quad\; (d) \qquad\quad\; (e) \qquad\quad\; (f) \qquad\quad\;\, (g) \qquad\quad\;\, (h) \qquad\quad\;\, (i) \qquad\quad\;\, (j)}
    \end{picture}
    
    \caption{Fused results (odd row) and error maps (even row) of different methods on the JiaXing dataset. (a) ground truth. (b) GSA. (c) SFIM. (d) Wavelet. (e) MTF\_GLP\_HPM. (f) CNMF. (g) Hyperpnn2. (h) HSpeSet2. (i) HyperKite. (j) Ours. The RGB image is shown by utilizing the 13-th, 30-th, and 64-th bands of the original HSI.}
    \label{fig:jiaxing}
\end{figure*}

\indent 4). Real dataset: This dataset comprises the LR-HSI and PAN images from the ZY-1 02D satellite. Their spatial resolution is 30 and 2.5m respectively. The spectral wavelength ranges from 395-2501nm with 166 bands. Due to the low SNR in some bands, we choose 76 bands in fusion experiments. According to the Wald protocol \cite{ref51}, we generate the downsampled LR-HSI and PAN images in the training stage and the corresponding reference HR-HSI is the original HSI. The last performance testing is on the original resolution's image pairs. Geometric registration between LR-HSI and PAN is done before the fusion process. Note that we upsample the LR-HSI by four and three times as in the part 2 of Fig. \ref{fig:network}.

The spatial patch size of these datasets are all cropped into 64×64, 64×64 and 4×4 for HR-HSI, PAN image and LR-HSI without overlap except for real dataset of 48×48, 48×48 and 4×4. We choose 80\% area of these datasets respectively in the training stage, and the rest is for performance testing. Gaussian random noise is added to the generated image with a zero mean and standard deviation of 0.01. 

These hyperspectral images owned different land types, from dense buildings to smoothed crops. Spectral characteristics vary dramatically across these datasets, making it pretty challenging to restore these coarse and fined details simultaneously. So the generality of the fusion method to various degrees of texture details is critical, and experimental results demonstrate the excellent generalization capability of the proposed network on these datasets.\\

\noindent\emph{B. Training Details}\\
\indent We train the proposed network using Adam optimizer for 500 epochs with the proposed loss function in section \uppercase\expandafter{\romannumeral3}-F. The batch size is fixed to 15, and the learning rate is set to 0.002 initially, then declines with the factor of 0.05 per 100 epochs. All experiments are run under the Paddle 2.0.2 framework and Python 3.7 environment on a single v100 graphic processing unit (GPU) with 16GB memory.\\
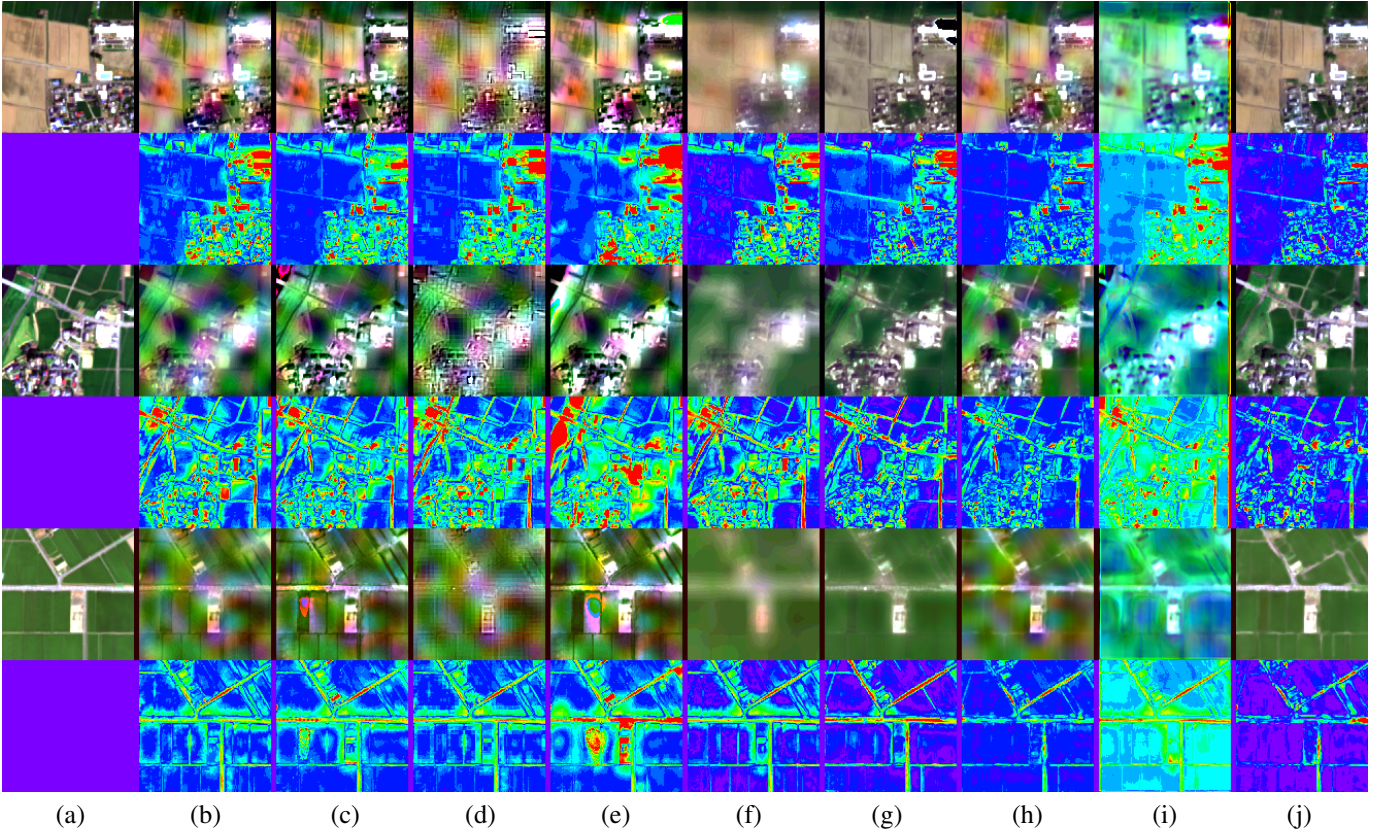
\begin{figure*}[t]
    \centering
    \includegraphics[width=1\textwidth, trim=0 0 0 0, clip]{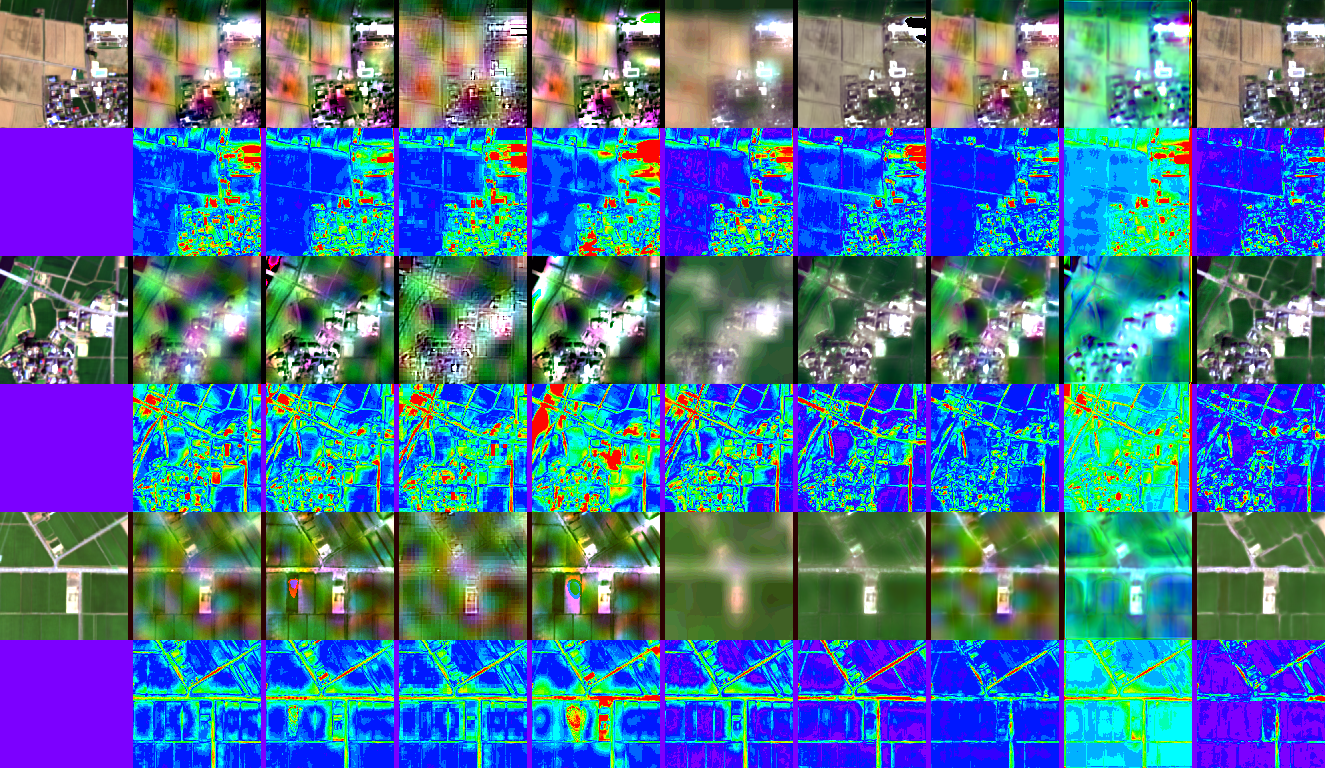}
    
    \begin{picture}(0,0)
    \put(-238,0){(a) \qquad\quad\; (b)  \qquad\quad\;\, (c) \qquad\quad\; (d) \qquad\quad\; (e) \qquad\quad\; (f) \qquad\quad\;\, (g) \qquad\quad\;\, (h) \qquad\quad\;\, (i) \qquad\quad\;\, (j)}
    \end{picture}
    
    \caption{Fused results (odd row) and error maps (even row) of different methods on the Chikusei dataset. (a) ground truth. (b) GSA. (c) SFIM. (d) Wavelet. (e) MTF\_GLP\_HPM. (f) CNMF. (g) Hyperpnn2. (h) HSpeSet2. (i) HyperKite. (j) Ours. The RGB image is shown by utilizing the 20-th, 40-th, and 60-th bands of the original HSI.}
    \label{fig:chikusei}
\end{figure*}

\linespread{1.35}   
\begin{table}[t]
    \centering
    \setlength{\abovecaptionskip}{0.cm}
    \caption{AVERAGE QUANTITATIVE RESULT ON JIAXING DATASET}
    \setlength{\tabcolsep}{1.5mm}{
    \begin{tabular}{  C{2cm} C{1cm} C{1cm} C{1cm} C{1cm} C{1cm}  }
\hline
Method&PSNR(↑)&SSIM(↑)&SAM(↓)&	ERGAS(↓)&SCC(↑)\\ 
\hline
GSA	&33.109&0.872&0.088&0.924&0.800\\
SFIM&33.422&0.881&0.087&0.898&0.818\\
Wavelet&31.666&0.825&0.090&1.039&0.713\\
MTF\_GLP\_HPM&31.409&0.873&0.090&1.079&0.770\\
CNMF&33.405&0.854&0.078&0.900&0.823\\
Hyperpnn2&\underline{35.179}&0.897&\underline{0.072}&\underline{0.766}&\underline{0.887} \\
HSpeSet2&34.911&\underline{0.904}&0.078&0.786&0.872 \\
HyperKite&29.312&0.846&0.139&1.363&	0.781\\
Ours&\bf{35.647}&\bf{0.905}&\bf{0.070}&\bf{0.736}&\bf{0.899} \\
\hline
    \end{tabular}}
    \label{tab:jiaxing}
\end{table}
\noindent\emph{C. Comparison Methods And Quality Measures Metrics}\\
\indent We compare the proposed Pgnet with several popular methods in the hyperpansharpening field from CS, MRA to VO and DL-based approaches. CS-based includes GSA, whereas MRA-based consist of SFIM, Wavelet and MTF\_GLP\_HPM. The VO-based method has CNMF, while selected Hyperpnn2, HSpeSet2 and HyperKite belong to DL-based methods. For a fair comparison, epochs and batch size of these DL-based methods are all adjusted to achieve their best performance, and other parameters are following the recommendation of the original work. In particular, considering the complexity of HyperKite, we reduce the batch size to 10 for memory limitation and follow the provided code setting that discards the DIP prior network.

\begin{figure*}[t]
    \centering
    \includegraphics[width=1\textwidth, trim=0 0 0 0, clip]{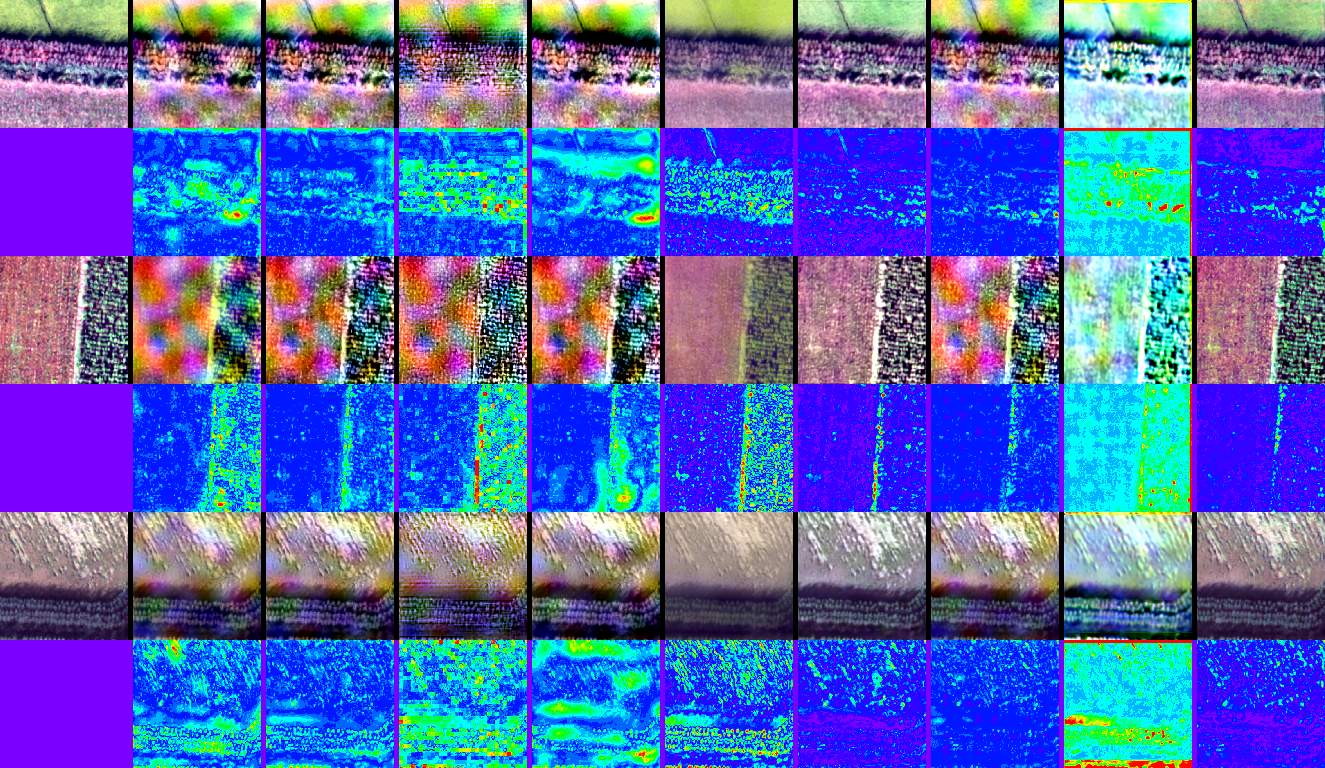}
    
    \begin{picture}(0,0)
    \put(-238,0){(a) \qquad\quad\; (b)  \qquad\quad\;\, (c) \qquad\quad\; (d) \qquad\quad\; (e) \qquad\quad\; (f) \qquad\quad\;\, (g) \qquad\quad\;\, (h) \qquad\quad\;\, (i) \qquad\quad\;\, (j)}
    \end{picture}
    
    \caption{Fused results (odd row) and error maps (even row) of different methods on the xiongan dataset. (a) ground truth. (b) GSA. (c) SFIM. (d) Wavelet. (e) MTF\_GLP\_HPM. (f) CNMF. (g) Hyperpnn2. (h) HSpeSet2. (i) HyperKite. (j) Ours. The RGB image is shown by utilizing the 36-th, 72-th, and 120-th bands of the original HSI.}
    \label{fig:xiongan}
\end{figure*}

\indent The quality indices used to measure and compare the performance of different methods include Peak Signal-to-Noise Ratio (PSNR) \cite{ref23}, structure similarity (SSIM) \cite{ref58}, Erreur Relative Globale Adimensionnelle de Synthèse (ERGAS) \cite{ref10}, Spectral Angle Mapper (SAM) \cite{ref10}, and spatial consistence coefficient (SCC) \cite{ref10}. PSNR and ERGAS measure the absolute errors between the result HR-HSI and reference images. SAM indicates angle distance, while SSIM and SCC represent the spatial and spectral similarity between fused and reference images. Among these metrics, the higher the PSNR, SSIM, SCC values, and the lower of other two indices, indicating the better fusion performance. In addition, the error map is also given to show the visual difference with the reference image in even rows of Fig. \ref{fig:jiaxing}, \ref{fig:chikusei} and \ref{fig:xiongan}. And the error map ranging from 0 to 0.05 represents the mean of all bands' absolute difference between the reference and fused results.\\
\linespread{1.35}   
\begin{table}[t]
    \centering
    \setlength{\abovecaptionskip}{0.cm}
    \caption{AVERAGE QUANTITATIVE RESULT ON CHIKUSEI DATASET}
    \setlength{\tabcolsep}{1.5mm}{
    \begin{tabular}{  C{2cm} C{1cm} C{1cm} C{1cm} C{1cm} C{1cm} }
\hline
Method&PSNR(↑)&SSIM(↑)&SAM(↓)&	ERGAS(↓)&SCC(↑)\\ 
\hline
GSA	&31.228&0.800&0.192&2.651&0.747\\
SFIM&31.678&0.832&0.198&4.471&0.778\\
Wavelet&31.022&0.796&0.192&3.005&0.718\\
MTF\_GLP\_HPM&28.176&0.809&0.209&7.301&0.685\\
CNMF&32.070&0.839&0.150&3.515&0.802\\
Hyperpnn2&34.879&\underline{0.903}&\underline{0.097}&\underline{2.009}&\underline{0.843} \\
HSpeSet2&\underline{35.103}&0.884&0.126&2.471&0.813 \\
HyperKite&30.704&0.731&0.233&5.545&0.626\\
Ours&\bf{36.266}&\bf{0.930}&\bf{0.080}&\bf{1.687}&\bf{0.889} \\
\hline
    \end{tabular}}
    \label{tab:chikusei}
\end{table}

\linespread{1.35}   
\begin{table}[t]
    \centering
    \setlength{\abovecaptionskip}{0.cm}
    \caption{AVERAGE QUANTITATIVE RESULT ON XIONGAN DATASET}
    \setlength{\tabcolsep}{1.5mm}{
    \begin{tabular}{  C{2cm} C{1cm} C{1cm} C{1cm} C{1cm} C{1cm} }
\hline
Method&PSNR(↑)&SSIM(↑)&SAM(↓)&	ERGAS(↓)&SCC(↑)\\ 
\hline
GSA&33.878&0.921&0.068&0.588&0.718\\ 
SFIM&35.499&0.947&0.066&0.519&0.810\\ 
Wavelet&32.904&0.869&0.075&0.711&0.710\\ 
MTF\_GLP\_HPM&33.557&0.942&0.067&0.635 &0.782\\ 
CNMF&34.193&0.885&0.046&0.517&0.809\\ 
Hyperpnn2&\underline{38.345}&0.965&\underline{0.041}&\underline{0.344}&\underline{0.907} \\
HSpeSet2&37.813&\underline{0.970}&0.055&0.405&0.839 \\
HyperKite&30.389&0.897&0.122&1.021&0.708\\
Ours&\bf{39.272}&\bf{0.971}&\bf{0.039}&\bf{0.327}&\bf{0.914} \\
\hline
    \end{tabular}}
    \label{tab:xiongan}
\end{table}
\noindent\emph{D. Experiments on Three Simulated Datasets}\\
\noindent\emph{1)	JiaXing Dataset}\\
\indent As shown in Fig. \ref{fig:jiaxing}, most of the existing fusion methods cause severely spectral and spatial distortion. They are unable to predict subtle details in the face of a huge ratio between two input images. In detail, CS and MRA-based methods all suffer from spectral ring artifacts when facing dense buildings, especially the Wavelet method. The results of CNMF have the blurring effect in the edge between buildings and roads, as shown in Fig. \ref{fig:jiaxing}(f). Differently, fused images of DL-based methods show a better visual effect than traditional methods. But they are still far from the ground truth except ours, such as the results of Hyperpnn2 and HyperKite have severe spectral distortion, which could also be seen from error maps. In contrast, as shown in Fig. \ref{fig:jiaxing}(j), the fused results of our method have few spectral differences with the ground truth and have a more detailed texture than other methods.

The average quantitative results in all test images also verify the capability of the proposed method in capturing the texture and edge details, as indicated by Table \ref{tab:jiaxing}. The best result in each column is highlighted in the bold style and underlined for the inferior result, which are the same styles as other tables. Compared with the secondary outcomes, the metric indices of ours improved by 1.33\% in PSNR, 0.11\% in SSIM, 2.78\% in SAM, 3.92\% in ERGAS and 1.35\% in SCC. Note that the HSpeSet2 achieves the secondary rank in four quantitative indices but shows poor visual performance in Fig. \ref{fig:jiaxing}. This phenomenon could attribute to the overlapping between grassland and green artifacts in Fig. \ref{fig:jiaxing}(a) and (h).

\noindent\emph{2)	Chikusei Dataset}\\
\indent Testing images in this dataset are generated from the bottom of the raw image, consisting of buildings and crops. Traditional fusion methods all induce severe spectral halo effect except for CNMF, as shown in Fig. \ref{fig:chikusei}(b)-(e). The error maps also show that the MTF\_GLP\_HPM method introduces the most difference with the reference image. Moreover, the results of CNMF and Hyperpnn2 suffer from blurring effects, especially around buildings and roads. In contrast, our fused results show excellent spectral and spatial fidelity, such as the edge between crops and buildings in Fig. \ref{fig:chikusei}(j). The error maps also indicate that there's little spectral distortion in the crop region.\\
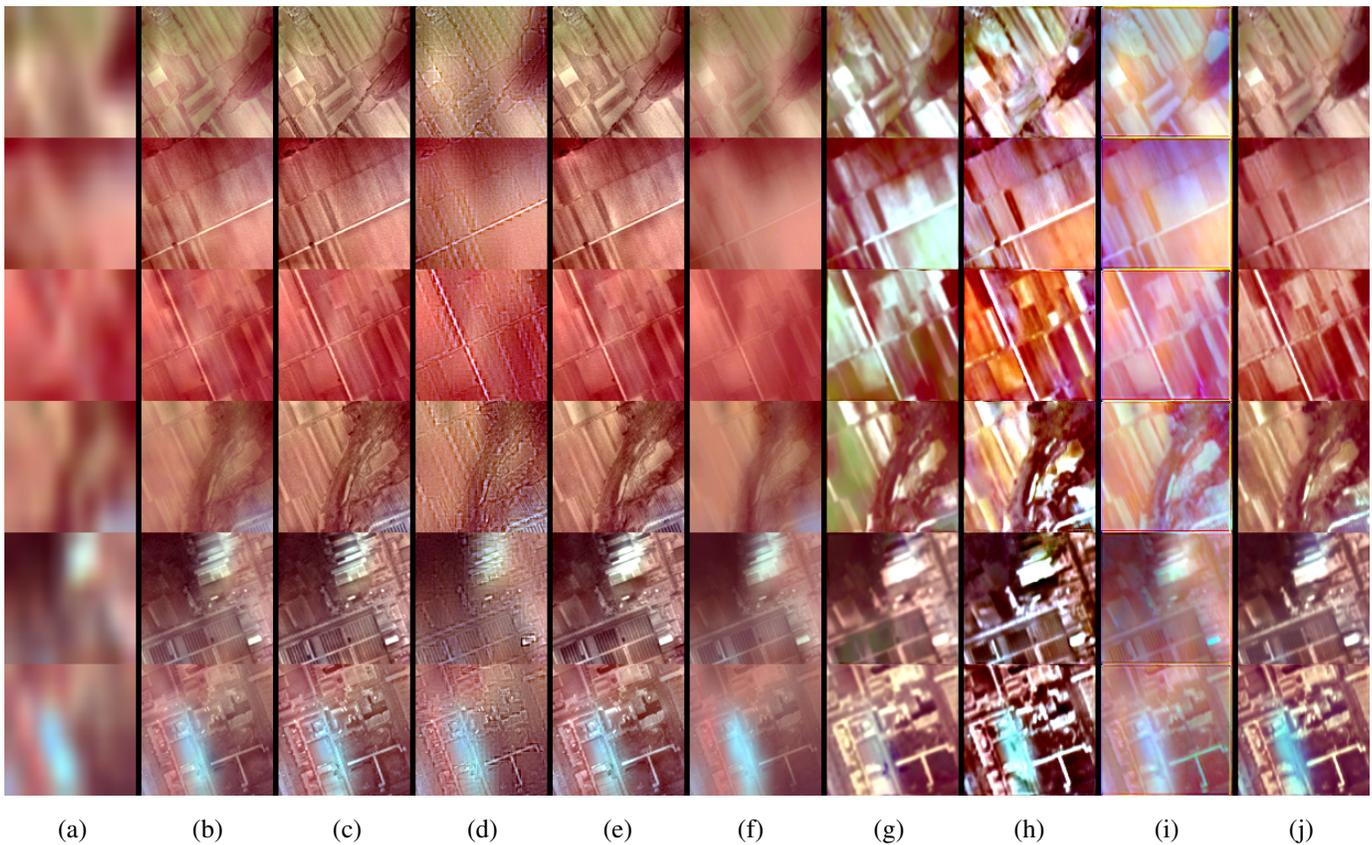
\begin{figure*}[t]
    \centering
    \includegraphics[width=1\textwidth, trim=0 0 0 0, clip]{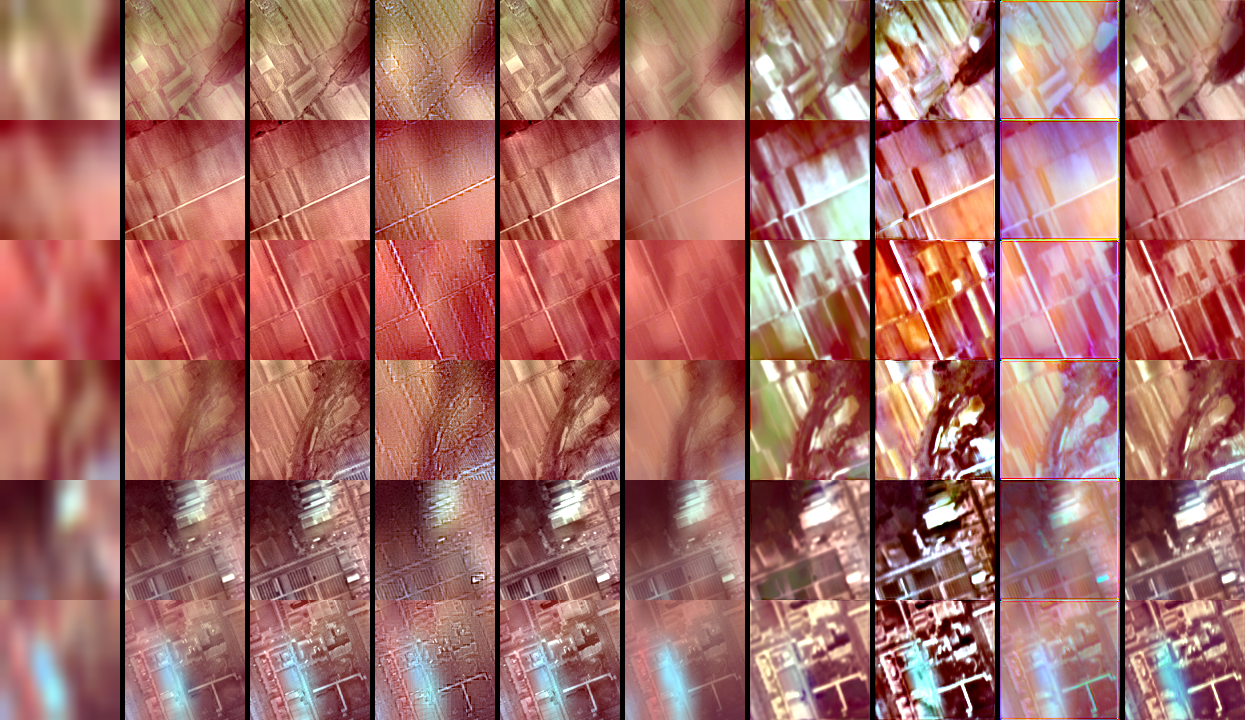}

    \begin{picture}(0,0)
    \put(-238,0){(a) \qquad\quad\; (b)  \qquad\quad\;\, (c) \qquad\quad\; (d) \qquad\quad\; (e) \qquad\quad\; (f) \qquad\quad\;\, (g) \qquad\quad\;\, (h) \qquad\quad\;\, (i) \qquad\quad\;\, (j)}
    \end{picture}
    
    \caption{Fused results of different methods on real dataset. (a) upsampled LR-HSI. (b) GSA. (c) SFIM. (d) Wavelet. (e) MTF\_GLP\_HPM. (f) CNMF. (g) Hyperpnn2. (h) HSpeSet2. (i) HyperKite. (j) Ours. The RGB image is shown by utilizing the 7-th, 16-th, and 35-th bands of the original HSI.}
    \label{fig:real}
\end{figure*}
\indent Quantitative results on Chikusei dataset also verify the superiority of Pgnet over other methods, as shown in Table \ref{tab:chikusei}. Advanced by a significant margin, the results of Pgnet improved by 3.31\%, 2.99\%, 17.53\%, 16.03\% and 5.46\% than inferior methods in five quality indices, respectively. The fourth method shows the worst index rank, which is consistent with visual analysis. Note that even CNMF achieves a better index rank than other traditional methods, its visual effect is worse than them in detail recovery, as shown in Fig. \ref{fig:chikusei}(f).\\

\noindent\emph{3)	Xiongan Datasets}\\
\indent The matiwan countryside in XiongAn is mainly occupied by crops, where has a pretty smooth texture except in the edge across different crops. Results of traditional methods and HSpeSet2 suffer from acutely spectral distortion, as shown in Fig. \ref{fig:xiongan}(b)-(e), (h), which might cause incorrect bias in quantitative retrieval of remotely sensed products like crop yields. The results of CNMF and Hyperpnn2 show great spectral fidelity, but they are still inferior to ours. As indicated by error maps, our method's results show better visual effects, especially in the first and third row of Fig. \ref{fig:xiongan}(j).

\indent Quantitative analysis was also implemented to provide a distinctive comparison for these methods. Due to the change of land types, these metrics all improve a certain degree for all methods. However, our method still achieves the best rank in five quality indices by a significant interval than other methods, as shown in Table \ref{tab:xiongan}. The above qualitative and quantitative results of our method demonstrate its effectiveness and generality on three datasets with coarse to fined texture and spectral details.\\

\noindent\emph{E. Experimental Results on Real Datasets	}\\
\indent Fusion results on the real dataset are shown in Fig. \ref{fig:real}. Due to the fused HR-HSI is under the same resolution as the original PAN image (2.5m), there's no ground truth to compare the fusion performance quantitatively. So we just show the upsampled LR-HSI in the first column of Fig. \ref{fig:real} for spectral fidelity comparison approximately.

As shown in Fig. \ref{fig:real}, the traditional methods show great texture details than DL-based methods except for Wavelet and CNMF. This phenomenon could attribute to the input image's large resolution gap between the training and testing stages in DL-based methods. However, not only do the traditional methods cost substantial computation time to fuse the HR-HSI as in Table \ref{tab:complexity}, but also, the spectral details restored in these methods may not be correct for there's no ground truth to guide the fusion process. Moreover, DL-based methods including Hyperpnn2, HSpeSet2 and HyperKite all suffer from severe spectral distortion compared with the upsampled LR-HSI. By comparison, the fused HR-HSI of our method shows the better bright contrast and spectral fidelity with upsampled LR-HSI, especially in the third, fourth and last rows of Fig. \ref{fig:real}(j). These restored spectral features could benefit accurate object recognition and segmentation.\\

\linespread{1.35}   
\begin{table}[t]
    \centering
    \setlength{\abovecaptionskip}{0.cm}
    \caption{AVERAGE QUANTITATIVE RESULTS ON CHIKUSEI DATASET WITH RATIO 8}
    \setlength{\tabcolsep}{1.5mm}{
    \begin{tabular}{  C{2cm} C{1cm} C{1cm} C{1cm} C{1cm} C{1cm} }
\hline
Method&PSNR(↑)&SSIM(↑)&SAM(↓)&	ERGAS(↓)&SCC(↑)\\ 
\hline
GSA&34.431&0.849&0.130&5.110&0.790\\ 
SFIM&31.562&0.842&0.159&16.630&0.698\\ 
Wavelet&34.020&0.834&0.128&5.433&0.772\\ 
MTF\_GLP\_HPM&29.248&0.819&0.166&20.129&0.663\\ 
CNMF&34.923&0.877&\underline{0.084}&3.818&0.849\\ 
Hyperpnn2&36.623&\underline{0.918}&0.092&\underline{3.670}&\underline{0.886} \\
HSpeSet2&\underline{37.157}&0.907&0.111&4.237&0.872 \\
HyperKite&34.764&0.852&0.174&9.884&0.781\\ 
Ours&\bf{37.561}&\bf{0.930}&\bf{0.079}&\bf{3.254}&\bf{0.917} \\
\hline
    \end{tabular}}
    \label{tab:ratio 8}
\end{table}

\linespread{1.35}   
\begin{table}[t]
    \centering
    \setlength{\abovecaptionskip}{0.cm}
    \caption{AVERAGE QUANTITATIVE RESULTS ON CHIKUSEI DATASET WITH RATIO 4}
    \setlength{\tabcolsep}{1.5mm}{
    \begin{tabular}{  C{2cm} C{1cm} C{1cm} C{1cm} C{1cm} C{1cm} }
\hline
Method&PSNR(↑)&SSIM(↑)&SAM(↓)&	ERGAS(↓)&SCC(↑)\\ 
\hline
GSA&37.281&0.895&0.110&8.508&0.868\\ 
SFIM&31.435&0.861&0.154&33.780&0.709\\ 
Wavelet&36.392&0.880&0.111&9.131&0.858\\ 
MTF\_GLP\_HPM&29.427&0.836&0.161&41.159&0.697\\ 
CNMF&38.235&0.930&\bf{0.061}&5.286&0.929\\ 
Hyperpnn2&\underline{40.461}&\underline{0.956}&0.074&\underline{5.060}&\underline{0.947} \\
HSpeSet2&39.661&0.936&0.094&6.818&0.917 \\
HyperKite&36.293&0.883&0.120&11.045&0.792 \\
Ours&\bf{41.204}&\bf{0.964}&\underline{0.066}&\bf{4.592}&\bf{0.956} \\
\hline
    \end{tabular}}
    \label{tab:ratio 4}
\end{table}

\noindent\emph{F.	Different Scales Experiments on Chikusei Dataset}\\
\indent To test the generality of the proposed fusion network on different fusion ratios, we choose the commonly used Chikusei dataset to implement this experiment. The ratios to be tested include 4 and 8, which are always the choices in current hyperpansharpening tasks. \\
\indent With the ratio decrease, as shown in Table \ref{tab:ratio 8} and \ref{tab:ratio 4}, these methods all score higher than in the ratio of 16 due to the mitigation of the fusion task's ill-posedness. However, the values of PSNR and SSIM indices achieved by our method are still higher than the secondary method by 1.80\%, 1.30\% and 1.84\%, 0.84\% in ratios 8 and 4, respectively. Although the first rank is achieved by the CNMF method in SAM index under ratio 4, our method performs better in the other four quality indices, demonstrating its excellent generality on different fusion scales.

\linespread{1.35}   
\begin{table}[t]
    \centering
    \setlength{\abovecaptionskip}{0.cm}
    \caption{AVERAGE QUANTITATIVE RESULT ON CHIKUSEI DATASETS WITH DIFFERENT ENDMEMBER NUMBER}
    \setlength{\tabcolsep}{1.5mm}{
    \begin{tabular}{  C{2cm} C{1cm} C{1cm} C{1cm} C{1cm} C{1cm} }
\hline
endmember number&PSNR(↑)&SSIM(↑)&SAM(↓)&	ERGAS(↓)&SCC(↑)\\ 
\hline
10&35.868&0.912&0.091&1.756&0.863 \\
20&36.266&0.930&0.080&1.687&0.889 \\
30&36.293&0.934&0.078&1.663&0.891 \\
40&36.317&	0.941&0.075&1.512&0.896\\ 
50&36.426&	0.942&	0.074&1.506&0.902\\ 

\hline
    \end{tabular}}
    \label{tab:endmember}
\end{table}

\linespread{1.35}   
\begin{table}[t]
    \centering
    \setlength{\abovecaptionskip}{0.cm}
    \caption{AVERAGE QUANTITATIVE RESULT ON CHIKUSEI DATASETS WITH DIFFERENT ATTENTION BLOCKS NUMBER}
    \setlength{\tabcolsep}{1.5mm}{
    \begin{tabular}{  C{2cm} C{1cm} C{1cm} C{1cm} C{1cm} C{1cm} }
\hline
attention blocks&PSNR(↑)&SSIM(↑)&SAM(↓)&	ERGAS(↓)&SCC(↑)\\ 
\hline
0&	35.227&	0.911&	0.096&	1.860&	0.861\\
1&	35.613&	0.917&	0.093&	1.819&	0.868\\
2&	35.738&	0.922&	0.091&	1.768&	0.873\\
3&	35.955&	0.925&	0.085&	1.723&	0.877\\
4&	36.266&0.930&	0.080&	1.687&	0.889\\
5&	36.356&	0.931&	0.078&	1.667&	0.892\\

\hline
    \end{tabular}}
    \label{tab:attention blocks}
\end{table}

\section{DISCUSSION}

This section first compares the fusion performance under the different number of endmembers and attention blocks. Then we test the different network designs include encoder, decoder and PDIN block in Pgnet to achieve the best fusion performance. Besides, we also discuss the effect of the proposed PDIN in injecting the PAN detail by depicting the transformation pattern of intermediate results. Then we test the different weights of SAM loss in Eq. (\ref{eq31}) and initialization methods of PDIN. Finally, we compare the model complexity of different methods, including inference time and parameters number.\\

\begin{figure*}[t]
    \centering
    \includegraphics[width=1\textwidth, trim=0 70 0 50, clip]{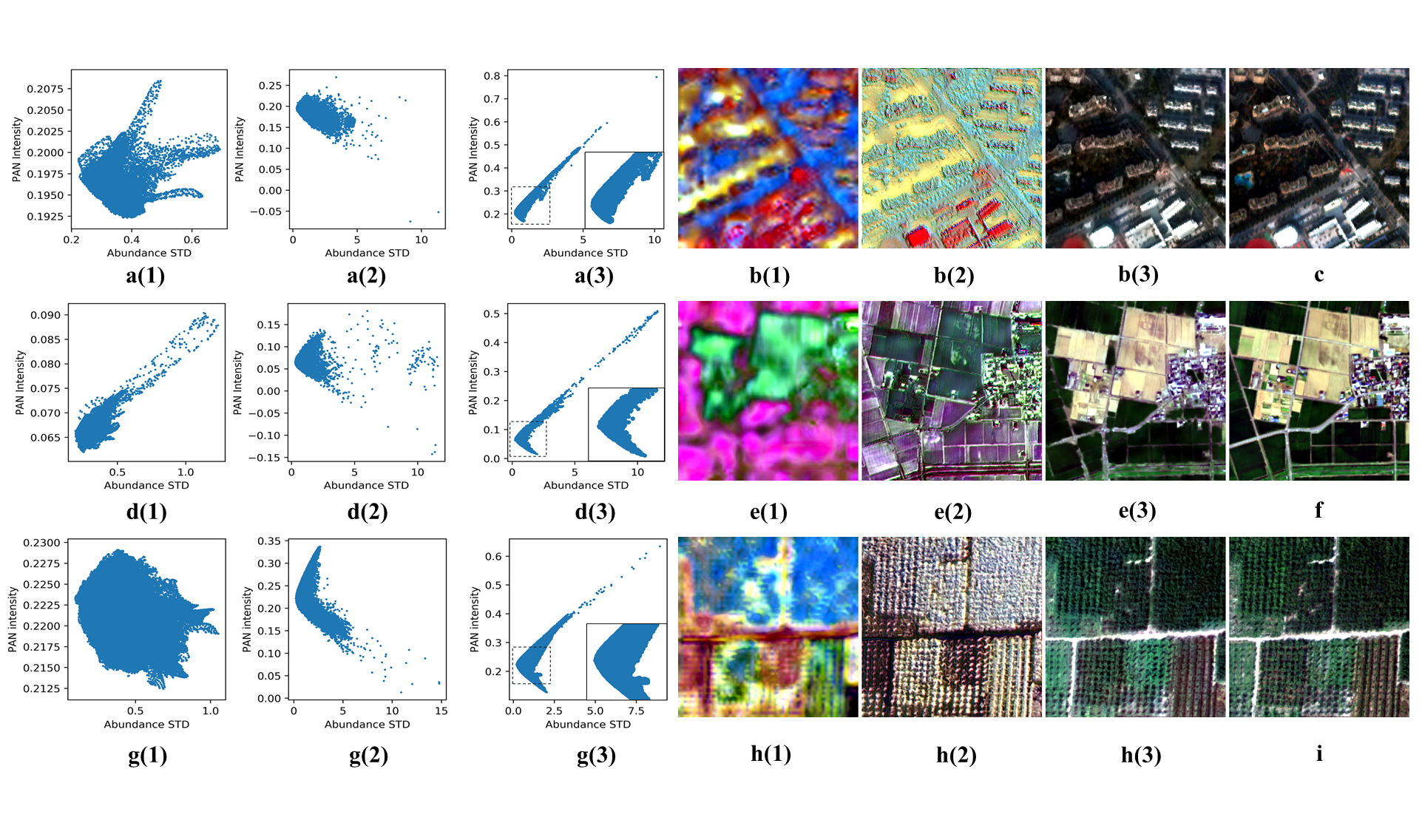}
    
    \caption{Intermediate results distribution and decoded images from the end of three network parts on three simulated datasets. Each row corresponds to one dataset. The first to the third columns are the data distribution (x-axis represent the Abundance STD and y-axis represent the PAN intensity) from the end of three network parts (the end of second upsample module, the end of second PDIN block in part 2 and the end of part 3 as in Fig. \ref{fig:network}). The fourth to sixth columns are the intermediate decoded images from above three network parts(by the last decoding operation directly). The last column is the reference image. i.e., In the first row, the scatter points of JiaXing dataset from three network parts are \textbf{a(1)}, \textbf{a(2)} and \textbf{a(3)}, and the decoded HSI from same network parts are \textbf{b(1)}, \textbf{b(2)} and \textbf{b(3)}, respectively. The last column-\textbf{c} is the reference JiaXing image. These notations are the same as Chikusei and Xiongan datasets in the second and third rows.}
    \label{fig:transform}
\end{figure*}

\linespread{1.35}   
\begin{table}[t]
    \centering
    \setlength{\abovecaptionskip}{0.cm}
    \caption{AVERAGE QUANTITATIVE RESULTS ON THE CHIKUSEI DATASET WITH DIFFERENT ENCODERS AND DECODERS. ‘en\_sig’ MEANS ENCODER WITH SINGLE CONVOLUTION BLOCK AND ‘de\_dou’ MEANS DECODER WITH DOUBLE CONVOLUTION BLOCKS AS IN THE SECTION \uppercase\expandafter{\romannumeral3}-E AND SAME FOR OTHER NOTATIONS.}
    \setlength{\tabcolsep}{1.5mm}{
    \begin{tabular}{  C{2cm} C{1cm} C{1cm} C{1cm} C{1cm} C{1cm} }
\hline
encoder/decoder&PSNR(↑)&SSIM(↑)&SAM(↓)&	ERGAS(↓)&SCC(↑)\\
\hline
en\_sig/de\_sig&32.647&	0.859&	0.130&	4.070&	0.748\\
en\_dou/de\_dou&\underline{35.760}&	\underline{0.920}&	\underline{0.091}&	\underline{1.775}&	\underline{0.875}\\
en\_dou/de\_sig&33.191&	0.863&	0.138&	3.819&	0.763\\
en\_sig/de\_dou&\bf{36.266}&	\bf{0.930}&	\bf{0.080}&	\bf{1.687}&	\bf{0.889}\\
\hline
    \end{tabular}}
    \label{tab:en-decoder}
\end{table} 

\noindent\emph{A.	Experiments on Endmembers and Attention Blocks Number.}\\
\indent To take a balance between network complexity and fusion performance, we test the different number of endmembers and attention blocks in the proposed Pgnet. Experiments on different situations are tested to determine the best compromise between performance and complexity. Its implements are on the commonly used and representative Chikusei dataset which is the same as other experiments in this section. Comparison results of endmember numbers from 10 to 50 on the Chikusei dataset are in Table \ref{tab:endmember}. This experiment is under the pre-defined self-attention blocks number of 4. 

With the increase of endmembers, the quality indices are inevitably improved, which is reasonable for the increment of parameters. However, due to the little increased margin of the quality indices when beyond 20 endmembers as shown in Table \ref{tab:endmember} and exponentially grown parameters, we choose 20 as the endmember baseline of the proposed network. Noticeably, even with ten endmembers, our method still achieves excellent results, verifying the validity of taking fusion operations in the low-dimensional abundance feature subspace.

Experiments on the different number of attention blocks from 0 to 5 are then done with the selected 20 endmembers. We set it to 4 in Pgnet for it achieved performance as in Table \ref{tab:attention blocks} and appropriate complexity.\\

\linespread{1.35}   
\begin{table}[t]
    \centering
    \setlength{\abovecaptionskip}{0.cm}
    \caption{EXPERIMENTS ON DIFFERENT COMBINATIONS OF THE PDIN. ‘$q_{weight}$’ IS THE GENERATED MOVING WEIGHT. 'PAN' MEANS THE LEARNED PAN WEIGHT—‘$P_1$’ AND ‘$P_2$’ IN Eq. (\ref{std_change_last}). THE NOTATION ‘bias’ REPRESENTS THE PROPOSED BIAS TERM IN Eq. (\ref{std_change_last}). ‘res’ MEANS THE RESIDUAL CONNECTION. THE TICK NOTATION MEANS THERE HAS THIS PART AND VICE VERSA.}
    \setlength{\tabcolsep}{1.5mm}{
    \begin{tabular}{  C{0.8cm} C{0.8cm} C{1cm} C{1cm} C{1cm} C{1cm} C{1cm} }

\hline
$q_{weight}$&PAN&PSNR(↑)&SSIM(↑)&SAM(↓)&	ERGAS(↓)&SCC(↑)\\ 
\hline
\XSolidBrush &	\XSolidBrush &32.380& 0.827& 0.099& 2.616& 0.697\\
\CheckmarkBold &	\XSolidBrush &	\underline{35.741}&	\underline{0.919}&	\underline{0.087}&	\underline{1.794}&	\underline{0.871}\\ 
\CheckmarkBold & \CheckmarkBold &	\bf{36.266}&	\bf{0.930}&	\bf{0.080}&	\bf{1.687}&	\bf{0.889}\\

\hline
\\
\hline
bias&res&PSNR(↑)&SSIM(↑)&SAM(↓)&	ERGAS(↓)&SCC(↑)\\ 
\hline
\XSolidBrush &	\XSolidBrush &36.174& 0.929& 0.083& 1.694& 0.888	\\
\CheckmarkBold &	\XSolidBrush &	\bf{36.282}&	\underline{0.930}&	\underline{0.081}&	\underline{1.688}&	\underline{0.888}\\
\CheckmarkBold & \CheckmarkBold &	\underline{36.266}&	\bf{0.930}&	\bf{0.080}&	\bf{1.687}&	\bf{0.889}\\

\hline
    \end{tabular}}
    \label{tab:PDIN}
\end{table}
\noindent\emph{B.	Experiments on Different Encoder and Decoder}

\indent To accurately extract the abundance feature and reconstruct the HR-HSI, we explore the appropriate encoder and decoder (en-decoder) design in this section. The main factors influencing the fusion performance include the number of CNN channels and blocks in the en-decoder. Here we test four kinds of en-decoder as in the first column of Table \ref{tab:en-decoder}. As shown in Table \ref{tab:en-decoder}, the best result is labeled with the bold style that corresponds to the fourth design. This en-decoder framework is more excellent than the first and third designs by a significant margin. Note that even compared to the second design with more parameters, the fourth en-decoder design still achieves better results in all quality indices, demonstrating the rationality and superiority of this kind of en-decoder design. \\

\noindent\emph{C. Experiments on Different PDIN Block }\\
\indent To verify the effectiveness of each part in PDIN, we successively remove these parts and compare the contributions of themselves, as shown in Table \ref{tab:PDIN}. Note that the experiments of the first table are equipped with ‘bias’ and 'res' terms. And the experiments of the second table are with ‘$q_{weight}$’ and ‘PAN’ in the first table. 

The first table in Table \ref{tab:PDIN} tests the importance of $q_{weight}$ and learned PAN weight in the fusion task. It indicates that the $q_{weight}$ contributes mostly to the achieved fusion performance by comparing the first two rows, i.e., the PSNR and SSIM indices improved by 10.38\% and 11.12\%, respectively. The addition of the learned PAN weight in Eq. (\ref{std_change_last}) also improves the fusion performance further in view of the last two rows of the first table. 

The second table in Table \ref{tab:PDIN} presents the contribution of the ‘bias’ term and residual connection in the PDIN block. It could be inferred that these two terms all improve the fusion performance. These improvements of fusion performance undoubtedly verify the effectiveness of all network parts in the proposed PDIN block.\\

\noindent\emph{D.	Transform Pattern Analysis of The Intermediate Results}\\
\indent In this section, to verify the rationality of the ‘fish’ distribution in Fig. \ref{fig:scatter points} and the effectiveness of the proposed PDIN in injecting the PAN detail, we depict the intermediate results' transformation pattern of the generated ‘HR-HSI’ during the forward propagation process. In detail, we output the abundance scatter points of all test images from the end of three selected network modules and the decoded HR-HSI from the same modules, as shown in Fig. \ref{fig:transform}. And the output abundance features are from the end of second upsample module, the end of second PDIN block in part 2 and the end of part 3 in Fig. \ref{fig:network}. These abundance maps are fed into the last decoding part to get the ‘HR-HSI’ of different parts (these HR-HSI correspond to the fourth—sixth column in Fig. \ref{fig:transform}). Then we multiply the HR-HSI by the SRF to get the PAN image and compute the abundance STD simultaneously. Lastly, we draw the data distribution of the PAN intensity and abundance STD in the first to the third column of Fig. \ref{fig:transform}.

These three datasets show nearly identical transform patterns from the first to the third column in Fig. \ref{fig:transform}. After upsampling to the same resolution with HR-HSI, the data distributions in the first column all have irregular shapes. The corresponding decoded HR-HSI in the fourth column which suffers from severe spatial and spectral distortion verifies the irrationality of data distributions in the first column. Then abundance details injected by the PDIN block change the data distribution into the second column of Fig. \ref{fig:transform}, where the distributions are partly same as the ‘fish’ distribution and the corresponding fusion performance improves a certain degree, as shown in the fifth column. Note that even we inject the pan details after the first upsample part through PDIN, the data distribution in the first column is still incorrect. It is because that the second upsampling operation of ratio 4 has just been accomplished and the first detail injection is under coarse resolution. 

Then the following attention part is used to restore the spatial and spectral details of HR-HSI to get completely enhanced HR-HSI, as shown in the sixth column of Fig. \ref{fig:transform}. And the data distributions in the third column become much similar to the ‘fish’ distribution after passing the successive attention blocks with PAN detail injection. Obviously, the reconstruction performance of the fused HR-HSI is greatly correlated with the correctness of the ‘fish’ distribution, which verifies the rationality of the ‘fish’ distribution. Also, this transformation patterns of intermediate results' distribution and achieved wonderful fusion performance demonstrate the effectiveness of the proposed PDIN in injecting the PAN details into the abundance feature by changing the data distribution.\\

\linespread{1.35}   
\begin{table}[t]
    \centering
    \setlength{\abovecaptionskip}{0.cm}
    \caption{EXPERIMENTS ON DIFFERENT SAM LOSS WEIGHT OF $\alpha$}
    \setlength{\tabcolsep}{1.5mm}{
    \begin{tabular}{ C{0.8cm} C{1cm} C{1cm} C{1cm} C{1cm} C{1cm} }
\hline
$\alpha$&PSNR(↑)&SSIM(↑)&SAM(↓)&ERGAS(↓)&SCC(↑)\\ 
\hline
0.001&\underline{35.776}&\underline{0.920}&0.094&\underline{1.777}&\underline{0.874}\\
0.01&\bf{36.266}&	\bf{0.930}&	\bf{0.080}&	\bf{1.687}&	\bf{0.889}\\
0.1&35.667&0.920&\underline{0.093}&1.839&0.871\\
1&34.802&0.910&0.098&2.766&0.849\\

\hline
    \end{tabular}}
    \label{tab:samweight}
\end{table} 

\linespread{1.35}   
\begin{table}[t]
    \centering
    \setlength{\abovecaptionskip}{0.cm}
    \caption{EXPERIMENTS ON DIFFERENT INITIALIZATION METHODS OF \\
    THE PDIN}
    \setlength{\tabcolsep}{1.5mm}{
    \begin{tabular}{ C{1cm} C{1cm} C{1cm} C{1cm} C{1cm} C{1cm} }
\hline
$\alpha$&PSNR(↑)&SSIM(↑)&SAM(↓)&ERGAS(↓)&SCC(↑)\\ 
\hline
Normal&\underline{36.405}&\underline{0.931}&0.080&\underline{1.669}&\underline{0.893}\\
Uniform&\bf{36.435}&\bf{0.932}&\underline{0.081}&\bf{1.659}&\bf{0.893}\\
Constant&36.266&0.930&\bf{0.080}&1.687&0.889\\
\hline
    \end{tabular}}
    \label{tab:PDINinit}
\end{table} 

\noindent\emph{E. Experiments on Different Parameters of $\alpha$ and Initialization Methods of PDIN}

\indent We test the different SAM loss weight—$\alpha$ in Eq. (\ref{eq31}) to achieve the best fusion performance. As shown in Table \ref{tab:samweight}, the weight of $0.01$ results in the best indices rank than other weights, which means it achieves the best balance between data fidelity and spectral constraint.\\
\indent To verify the generality of the proposed PDIN with different initialization parameters, we take experiments with different initialization methods, including Normal random, Uniform random and Constant initialization, as shown in Table \ref{tab:PDINinit}. Note that the Normal initialization is of mean value $0$ and STD value $1$, while Uniform initialization is sampled evenly between $-1$ and $1$. The Constant initialization is inspired by Fig. \ref{fig:scatter points} that two slopes $k_1$, $k_2$ are initialized with $1$ and $-1$ while the intercepts—$b_1$ and $b_2$ are initialized with the same value of $0.3$ as in Eq. (\ref{redline1}) and (\ref{redline2}). As shown in Table \ref{tab:PDINinit}, different initialization methods all achieve excellent fusion performance, which demonstrates the robustness of the proposed PDIN. And we choose the Constant initialization method considering its fast convergence rate and achieved comparable fusion performance.\\

\linespread{1.35}   
\begin{table}[t]
    \centering
    \setlength{\abovecaptionskip}{0.cm}
    \caption{AVERAGE TESTING TIME AND MODEL PARAMETERS ON CHIKUSEI DATASET}
    \setlength{\tabcolsep}{1.5mm}{
    \begin{tabular}{  C{2cm} C{1.5cm} C{2cm}}
\hline
Method&	Time(s)(↓)&	Parameters(m)(↓)\\
GSA	&4.7202&	—\\
SFIM&	17.9695&	—\\
Wavelet&	4.3358&	—\\
MTF\_GLP\_HPM&	12.5180&	—\\
CNMF&	43.9069&	—\\
Hyperpnn2&	\bf{0.0048}&	0.13\\
HSpeSet2&	\underline{0.0064}&	\underline{0.11}\\
HyperKite&	0.1393&	0.55\\
Ours&	0.0512&	\bf{0.049}\\

\hline
    \end{tabular}}
    \label{tab:complexity}
\end{table}

\noindent\emph{F. Comparison of Computational Complexity}\\
\indent We compare the average inference time and parameters number of different methods on the Chikusei dataset with the patch size of 128×256×256, as listed in Table \ref{tab:complexity}. Traditional methods all suffer from great time costs like CNMF and SFIM, which spend about 44 and 18 seconds to fuse one image patch, respectively. In contrast, DL-based methods all have instant inference time. For example, Hyperpnn2 and HSpeSet2 take 0.0048 and 0.0064 seconds to predict HR-HSI, respectively. Through a short inference time of these two methods, our method could get better fusion performance and generality as in section \uppercase\expandafter{\romannumeral4}. Moreover, the parameter number of ours (only 0.049 million) and the marginal increment of inference time—0.0464 than Hyperpnn2 imply that our method can better trade-off between fusion performance and model complexity.

\section{CONCLUSION}
\indent In this paper, a lightweight unmixing-based pan-guided fusion network is proposed to release the ill-posed hyperpansharpening task when facing the extremely large fusion ratio and hundreds of bands of the HSI. Moreover, inspired by the linear and nonlinear relationships between PAN intensity and abundance STD, we propose an interpretable PAN detail inject network (PDIN), which injects the high-frequency detail from PAN image into the low-dimensional abundance feature. Experimental results on simulated and real datasets demonstrate the superiority and generality of the proposed fusion framework and PDIN block. The intermediate results of testing images on the different network parts also verify the rationality of the 'fish' distribution and the effectiveness of the proposed PDIN in injecting the PAN detail by correcting the data distribution. In addition, this fusion network could be adapted to other ratios' fusion tasks for it achieved excellent generalization performance on different fusion ratios' experiments. Although our method obtained excellent fusion performance, the generality of the already trained model on different hyperspectral datasets with different spectral bands still needs further research.\\

\noindent\uppercase{ACKNOWLEDGEMENT}

The authors would like to thank Baidu AI Studio for providing the computing power supports.




\begin{IEEEbiography}
[{\includegraphics[width=1in,height=1.25in,clip,keepaspectratio]{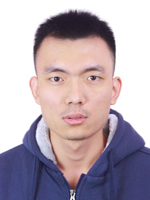}}] 
{Shuangliang Li} received the B.E. degree in Geographical Science from Hubei University, Wuhan, China, in 2019. He is now pursuing M.S. degree in Photogrammetry and Remote Sensing from China University of Geosciences, Wuhan, China.
His research interests include hyperspectral image processing, image fusion, and deep learning.
\end{IEEEbiography}
\vspace{-5 mm}

\begin{IEEEbiography}
[{\includegraphics[width=1in,height=1.25in,clip,keepaspectratio]{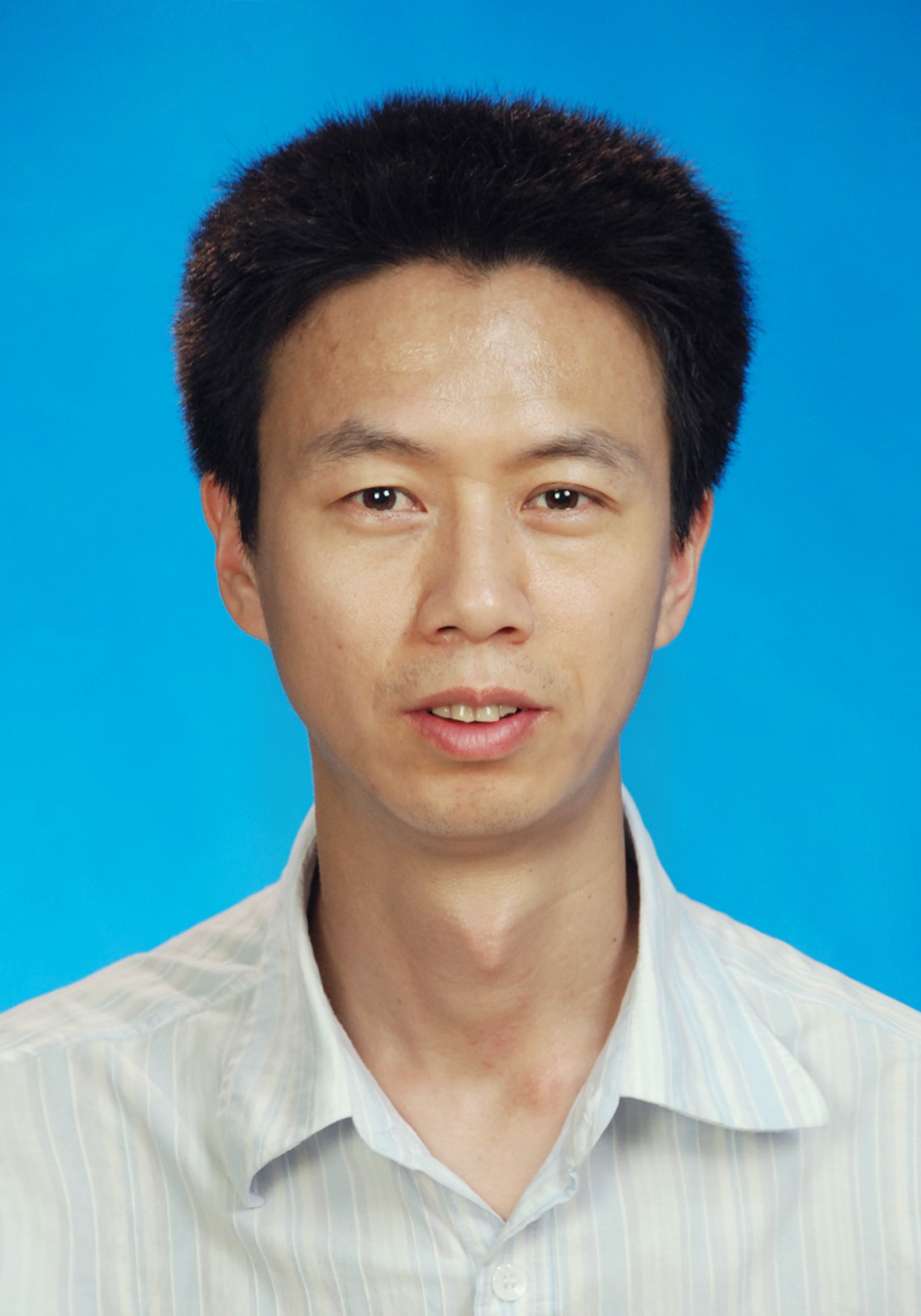}}] 
{Yugang Tian} received the B.E. degree in Surveying and mapping engineering from Wuhan University, Wuhan, China, in 2000, and the M.S. degree in geodesy and geomatics engineering from Wuhan University, Wuhan, China, in 2003. He received the Ph.D. degree in physical Geography from Beijing Normal University, Beijing, China, in 2006.

Since 2009, he has been an Associate Professor with the School of Geography and Information Engineering, China University of Geosciences, Wuhan, China. From 2014 to 2015, he was a visiting scholar in the Department of Civil and Environmental Engineering at Cornell University in the United States. His research interests include urban and environmental monitoring, image processing, and pattern cognition.

\end{IEEEbiography}
\vspace{-5 mm}

\begin{IEEEbiography}
[{\includegraphics[width=1in,height=1.25in,clip,keepaspectratio]{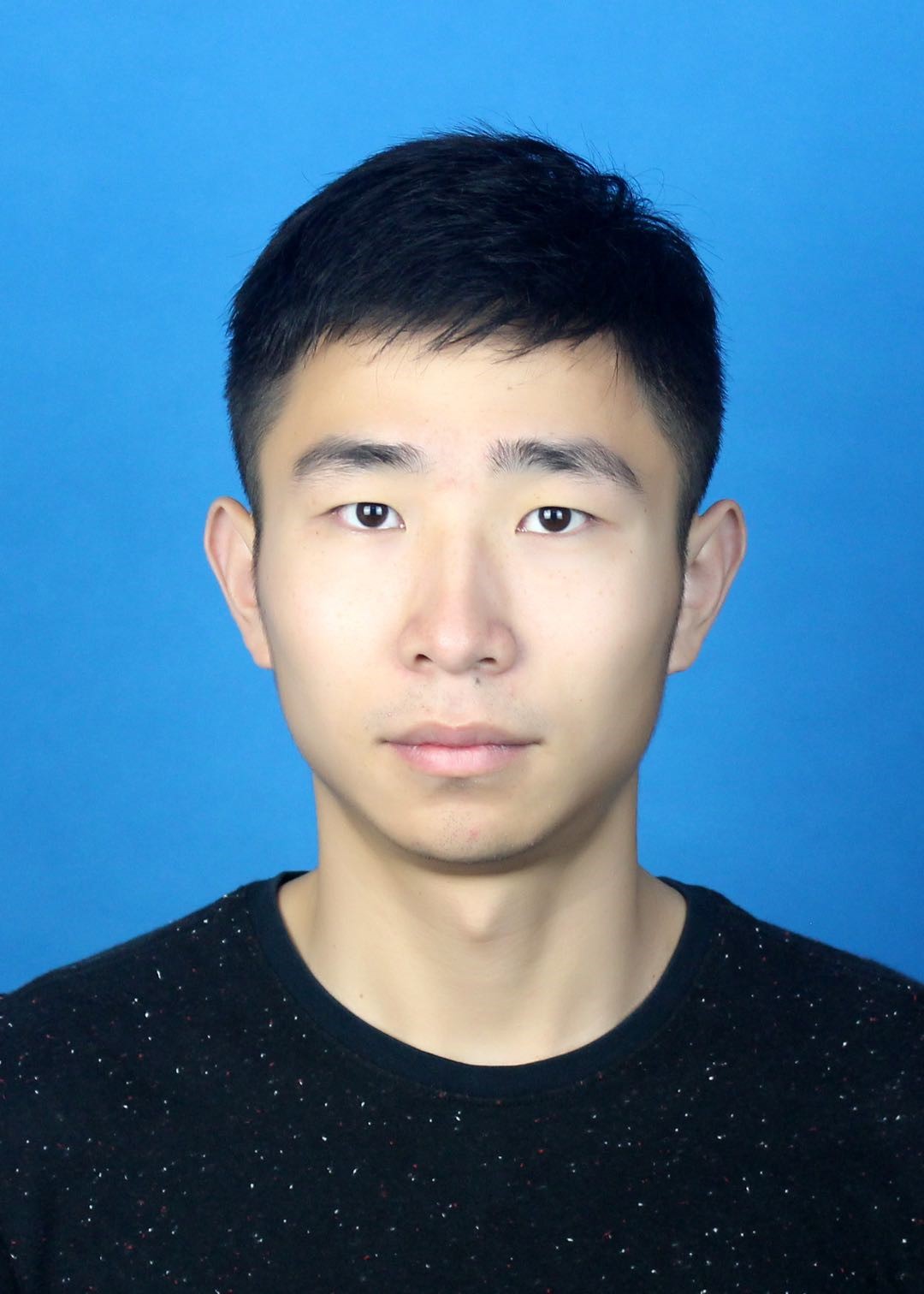}}] 
{Hao Xia} received the B.E. degree in Remote sensing science and technology from China University of Geosciences, Wuhan, China, in 2019, where he is currently pursuing the M.S. degree in Photogrammetry and remote sensing. His research interests include remote sensing image analysis and change detection.
\end{IEEEbiography}
\vspace{-5 mm}

\begin{IEEEbiography}
[{\includegraphics[width=1in,height=1.25in,clip,keepaspectratio]{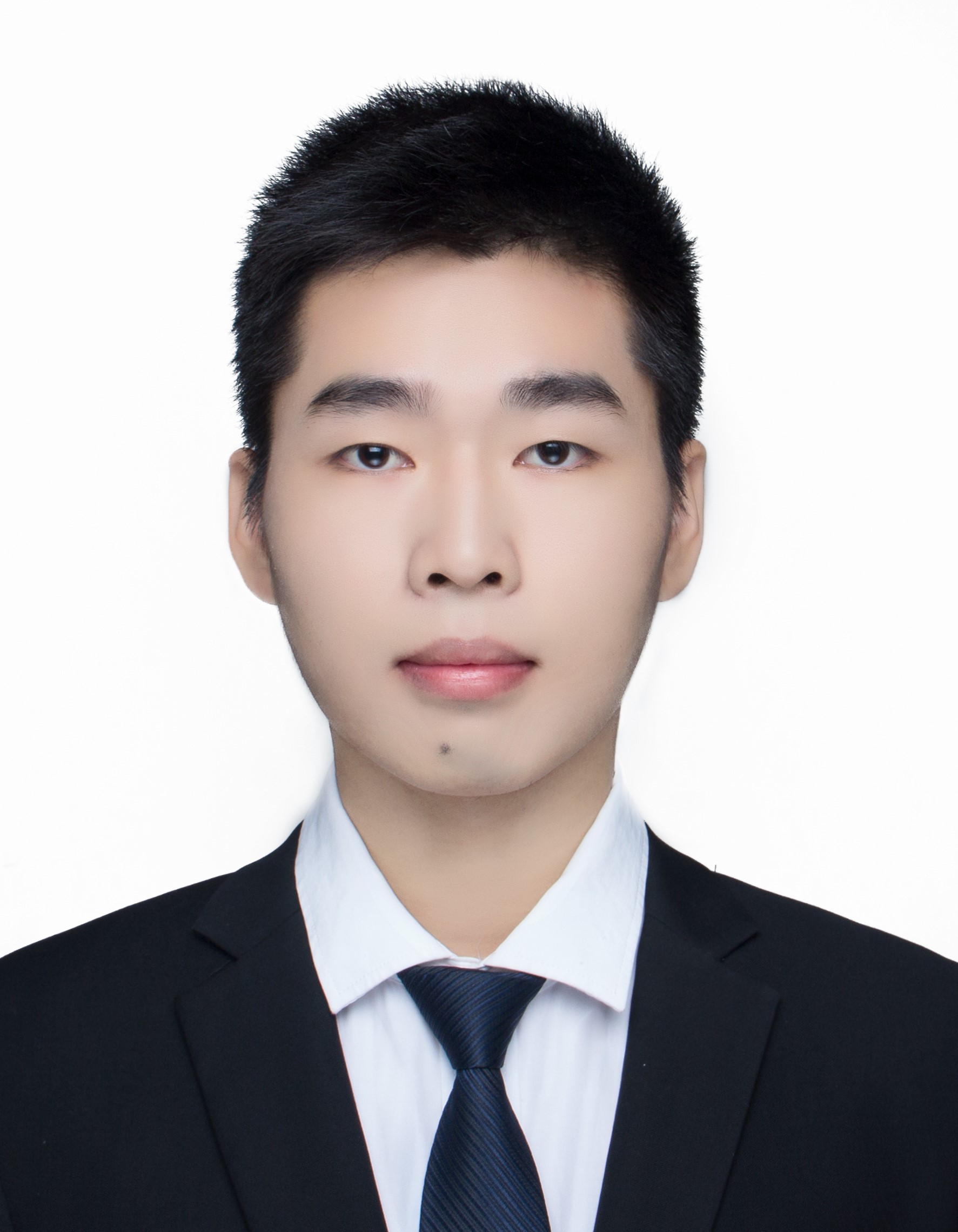}}] 
{Qingwei Liu} received the bachelor’s degree in Remote sensing science and technology from China University of Geosciences, Wuhan, China, in 2019, where he is currently pursuing the M.S. degree in Photogrammetry and remote sensing. His research interests include remote sensing image processing and computer vision.
\end{IEEEbiography}

\end{document}